\documentclass[12pt]{article}
\usepackage{times}
\usepackage{graphicx,amsmath,amssymb,color,braket}
\usepackage{float}
\usepackage[colorlinks]{hyperref}
\usepackage{cite}

\newenvironment{sciabstract}{%
	\addtolength\leftmargini{-0.2in}
	\begin{quote} \bf}
	{\end{quote}}
\makeatletter
\renewcommand\@biblabel[1]{#1.}
\makeatother

\title{Quantum interference enables constant-time\\ quantum information processing\\[0.2cm]
\large Single-step Quantum Kravchuk Transform}

\author{%
	M.~Stobi\'nska$^1$,
	A.~Buraczewski$^1$,
	M.~Moore$^2$,
	W.~R.~Clements$^2$,\\
	J.~J.~Renema$^3$,
	S.~W.~Nam$^4$,
	T.~Gerrits$^4$,
	A.~Lita$^4$,\\
	W.~S.~Kolthammer$^2$,
	A.~Eckstein$^2$,
	I.~A.~Walmsley$^2$\\
	\\
	\normalsize{$^1$Institute of Theoretical Physics, University of Warsaw,}\\
	\normalsize{ul.\ Pasteura 5, 02-093 Warsaw, Poland,}\\
    \normalsize{$^2$Clarendon Laboratory, University of Oxford,}\\
    \normalsize{Parks Road, Oxford OX1 3PU, United Kingdom,}\\
    \normalsize{$^3$Complex Photonic Systems (COPS), MESA+ Institute for Nanotechnology}\\
    \normalsize{University of Twente, P.O. Box 217, 7500 AE Enschede, Netherlands,}\\
    \normalsize{$^4$National Institute of Standards and Technology,}\\
    \normalsize{325 Broadway, Boulder, CO 80305, USA$^{\ast}$}\\
    \\
	\normalsize{$^\ast$Contribution of NIST, an agency of the U.S. government, not subject to copyright.}
}

\date{}

\begin{document}

\baselineskip24pt
\maketitle

	\begin{sciabstract}\baselineskip16pt
	It is an open question how fast information processing can be performed and whether quantum effects can speed up the best existing solutions. Signal extraction, analysis and compression in diagnostics, astronomy, chemistry and broadcasting builds on the discrete Fourier transform.  It is implemented with the Fast Fourier Transform (FFT) algorithm that assumes a periodic input of specific lengths, which rarely holds true. A less-known transform, the Kravchuk-Fourier (KT), allows one to operate on finite strings of arbitrary length. It is of high demand in digital image processing and computer vision, but features a prohibitive runtime. Here, we report a one-step computation of a fractional quantum KT. A quantum $d$-nary (qudit) architecture we use comprises only one gate and offers processing time independent of the input size. The gate may employ a multiphoton Hong-Ou-Mandel effect. Existing quantum technologies may scale it up towards diverse applications.
	\end{sciabstract}

    \noindent
    \textbf{MAIN TEXT} 

	\section*{Introduction}
	\vskip-3mm

	Science, medicine and engineering demand efficient information processing. It is a long-standing goal to use quantum mechanics to significantly improve such computations\cite{Zoller2005}. The processing routinely involves examining data as a function of complementary variables, e.g., time and frequency. This is done by the Fourier transform approximations which accurately compute inputs of $2^n$ samples in $O(n 2^n)$ steps\cite{Brigham1988}. In the quantum domain, an analogous process exists, namely a Fourier transform of quantum amplitudes\cite{Nielsen2000}, which requires exponentially fewer $O(n \log n)$ quantum gates. Here, we report a quantum fractional Kravchuk-Fourier transform, a related process suited to finite string processing\cite{Atakishiyev1997}. Unlike previous demonstrations\cite{Weimann2015,Crespi2016}, our architecture involves only one gate, resulting in constant-time processing of quantum information. The gate exploits a generalized Hong--Ou--Mandel effect\cite{Hong1987}, the basis for quantum-photonic information applications\cite{Makino2016}. We perform a proof-of-concept experiment by creation of large photon number states, interfering them on a beam splitter and using photon-counting detection.
	
	The discrete Fourier transform (DFT) is an efficient approximation to the Fourier transform (FT). The signal $(x_0, x_1, \ldots, x_S)$ is taken to be samples of one period of a continuous function, and is turned into a new sequence $(X_0, X_1, \ldots, X_S)$ where
	\begin{equation}
	\mathrm{X}_k = \frac{1}{\sqrt{S+1}}\sum_{l=0}^{S} e^{-i 2\pi \tfrac{ k l }{S+1}} \cdot x_l, \quad k = 0, \ldots, S.
	\label{eq:DFT}
	\end{equation}

	The DFT does not, however, reproduce all essential features of the FT. In some cases, a transform which is a fractional power of the FT, the $\alpha$-fractional FT where $0 \!\le \alpha \le \!1$, yields advantages\cite{Sejdic2010}. For $\alpha = 0$ this transform is the identity, while for $\alpha = 1$ this is the FT. If $\alpha=1/n$, where $n = 2,3,4,\ldots$, a composition of $n$ $\alpha$-fractional FTs amounts to the FT. This intuitive property does not hold true for the $\alpha$-fractional DFT (Supplementary Materials, SM), which generalizes the DFT but for $\alpha=1$ it reduces to Eq.~\ref{eq:DFT}. This is because the $\alpha$-fractional FT can be seen as a circular rotation of the signal in the time-frequency plane by angle $\frac{\pi\alpha}{2}$, while the $\alpha$-fractional DFT is an elliptical rotation in this plane which requires additional computation steps to properly approximate the $\alpha$-fractional FT\cite{Sejdic2010}.

\begin{figure}[p]\centering
	\includegraphics[height=4cm]{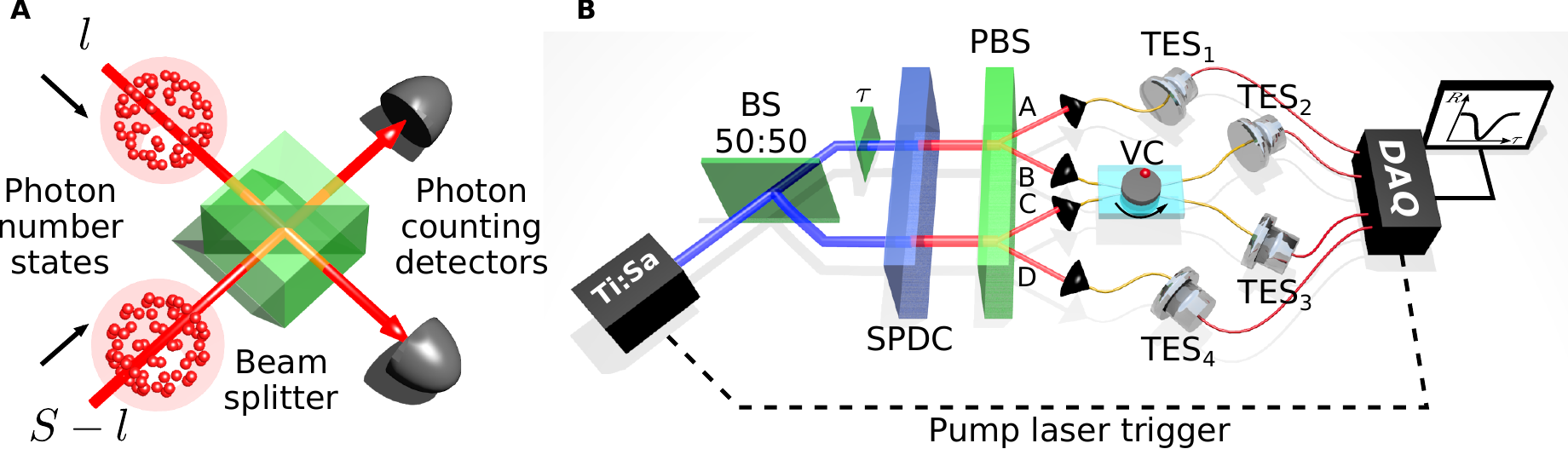}
	\caption{\textbf{Photonic implementation of a fractional QKT.} (\textbf{A}) HOM interference of photon number states on a variable beam splitter followed by two photon counting detectors, (\textbf{B}) Setup: Ti:Sa -- titanium-sapphire laser pump (blue), BS -- $50:50$ beam splitter, $\tau$ -- optical phase delay, SPDC -- periodically-poled potassium titanyl phosphate (PP-KTP) nonlinear spontaneous parametric down conversion waveguide chip which produces photon number correlated states (red), PBS -- polarization beam splitter, VC -- variable coupler, TES -- transition edge sensors, DAQ -- data acquisition unit.}
	\label{BS}
\end{figure}

\begin{figure}[p]\centering
	\includegraphics[height=7cm]{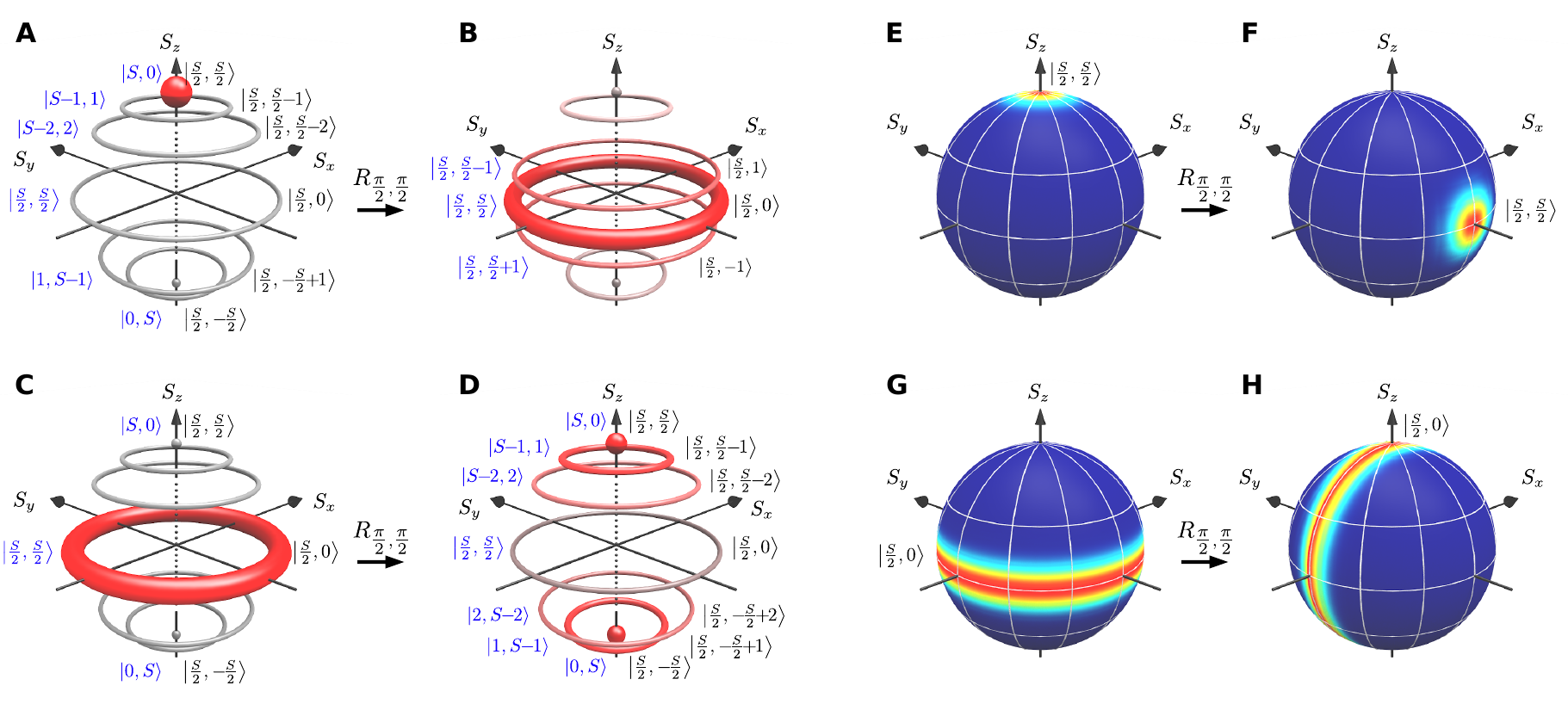}
	\caption{\textbf{HOM interference and QKT on a Bloch sphere.} (\textbf{A}--\textbf{D}) Two-mode Fock states (blue) correspond to Dicke states (black) -- the basis of spin $\tfrac{S}{2}$ states. HOM interference turns Dicke states into a superposition of them. This coincides with a rotation $R_{\theta,\varphi}$ in the Dicke state basis. The two most distinct cases are shown: the rotation $R_{\tfrac{\pi}{2},\tfrac{\pi}{2}}$ of the pole $\ket{\tfrac{S}{2},\tfrac{S}{2}}$ and of the great circle state $\ket{\tfrac{S}{2},0}$. (\textbf{E}--\textbf{H}) Q-function representation of a-d. HOM interference implements a rotation on the Bloch sphere by $\theta=\tfrac{\pi}{2}$ around $S_{x}$ of input $S_z$-eigenbasis Dicke states and thus, the full QKT, cf.\ Eq.~\ref{eq:KT}. The sequence $(x_0, x_1, \dots, x_S)$ is $(1,0, 0, \dots, 0)$ in (\textbf{A}) and $(0, \dots, 1, \dots, 0)$ in (\textbf{C}). The QKT transfers the input -- a position eigenstate -- into the same state but in $S_y$ basis -- a momentum eigenstate. }
	\label{Bloch}
\end{figure}

    The DFT is powerful due to the fast Fourier transform algorithm (FFT)\cite{Brigham1988}. Using an FFT lowers the number of operations from $O(2^{2n})$ to $O(n2^n)$ which nevertheless remains a bottleneck in signal processing\cite{Morgenstern1973}. The FFT employs a ``divide and conquer'' method to recursively split Eq.~\ref{eq:DFT} into $2^n$ sums which can be processed quickly, and therefore is applicable to signals of period $2^n$. Notably, the minimal number of operations required to implement the DFT is unknown\cite{Ailon2015}. The quantum Fourier transform (QFT), the cornerstone of quantum algorithms\cite{Klappenecker,Gottesman}, enables implementation of the DFT on quantum amplitudes with $O(n \log n)$ operations by processing $n$ qubits ($n$ quantum bits encode $2^n$ amplitudes)\cite{Hales2000}.  

    In many applications, e.g.\ bioimaging, the signals are typically not periodic and are random in length. For such cases, the Kravchuk transform (KT) is a useful alternative to the FFT because it can be applied to finite signal processing\cite{Yap2003,Kumar2015}. The KT computes orthogonal moments corresponding to the Kravchuk polynomials, which are discrete and orthogonal with respect to a binomial distribution in the data space\cite{Atakishiyev1997}. By varying a parameter of the binomial distribution, one is able to set the fractionality $\alpha$ of the KT (Supplementary Materials, SM). This feature allows to explore a specific region of interest of an image. To illustrate the action of a KT, the numerical study in Fig.~\ref{brains} in the SM demonstrates advantages of the KT over FFT in reconstructing test images.
 
    The KT's computational time is equal to the DFT's runtime\cite{Venkataramana2011} (SM) and implementations with lower number of operations are of high demand. Recently, quantum KTs (QKTs) have been realized in waveguides with two photons, but they are difficult to scale up and their fractionality is fixed by waveguide length\cite{Weimann2015,Crespi2016}.       
         
   	The $\alpha$-fractional KT employs the weighted Kravchuk polynomials $\phi_k^{(p)}(q,S)$\cite{Atakishiyev1997} which are real-valued and correspond to wave functions of finite harmonic oscillators
   	\begin{equation}
	\mathrm{X}_k= \sum_{l=0}^{S} e^{-i \tfrac{\pi\alpha}{2}\tfrac{S}{2}}\, 
	e^{i \tfrac{\pi}{2}(l-k)} \, \phi^{(p)}_k(l - S p,S) \cdot x_l, \quad k = 0, \ldots, S,
	\label{eq:KT}
	\end{equation}
	where $p=\sin^2 \left(\tfrac{\pi \alpha}{4}\right)$. Unlike plane waves, $e^{-i 2\pi \tfrac{ k l }{S+1}}$, the polynomials are defined and orthogonal on a set of $S+1$ points. This enables one to transform the signal as a finite sequence rather than as an infinite periodic one. In the limit of $S \to \infty$, $\phi_k^{(p)}(q,S)$ tend to  eigenfunctions of quantum harmonic oscillators and the $\alpha$-fractional KT reproduces the $\alpha$-fractional FT. Eq.~\ref{eq:KT} can be viewed in terms of overlaps of two spin $S/2$ states, in which they are prepared as eigenstates of $S_3$ and one undergoes a rotation by angle $\tfrac{\pi\alpha}{2}$ generated by $S_1$, $e^{i \tfrac{\pi}{2}(l-k)}\phi^{(p)}_k(l - S p,S) =\break \langle \tfrac{S}{2}; \tfrac{S}{2} - k \rvert e^{i \tfrac{\pi\alpha}{2} S_1} \vert \tfrac{S}{2}; \tfrac{S}{2} - l \rangle$.	
		
\section*{Results}
\vskip-3mm
		
	In this Report, we demonstrate a single-step QKT with tunable fractionality using quantum effects, based on multi-particle bosonic interference resulting from an exchange interaction. To this end, we interfere photon number states (light pulses with definite particle number) on a beam splitter (BS) with an adjustable splitting ratio. This leads to a multi-particle Hong--Ou--Mandel (HOM) effect\cite{Campos1989} which we observe for states with up to five photons. This QKT implementation enables constant-time quantum information processing for qudit data encoding which is set by the total number of interfering particles $S$, allowing up to $d\!=\!S\!+\!1$ signal samples. 
	
	Photon number (Fock) states $\ket{l}\! =\! \tfrac{(a^{\dagger})^{l}}{\sqrt{l!}} \ket{0}$ and $\ket{S-l}\! = \! \tfrac{(b^{\dagger})^{S-l}}{\sqrt{(S-l)!}} \ket{0}$ impinging on a beam splitter (BS) exhibit a generalized HOM effect, Fig.~\ref{BS}(\textbf{A}). A BS interaction between two such inputs described by annihilation operators $a$ and $b$ is $U_{BS}=\exp\{\tfrac{\theta}{2} (a^{\dagger}b e^{-i \varphi} - ab^{\dagger} e^{i \varphi})\}$, where $r=\sin^2\tfrac{\theta}{2}$ is the BS reflectivity (defined as the probability of reflection of a single photon) and $\varphi$ is the phase difference between the reflected and transmitted fields\cite{Kim2002}. Since $\varphi$ does not influence our experiments, we assume $\varphi = \tfrac{\pi}{2}$ for convenience. If the BS is balanced ($r=0.5$), two photons at the input ports will leave through the same exit port. This is known as photon bunching\cite{Hong1987}. Similar effects hold for multiphoton number states\cite{Campos1989}. This is reflected in the probability amplitudes of detecting $\ket{k}$ and $\ket{S-k}$ behind the BS, $\mathcal{A}^{(r)}_S(k,l) =\break e^{-i \theta \tfrac{S}{2}} \, \langle k, S-k \lvert U_{BS} \rvert l, S-l \rangle$. This is important for implementing the KT, since $\mathcal{A}^{(r)}_S(k,l) = e^{-i \theta \tfrac{S}{2}} \, e^{i \tfrac{\pi}{2}(l-k)} \cdot \phi^{(r)}_k(l - S r,S)$; thus, if we send a quantum state $\ket{\Psi} = \sum_{l=0}^S x_l \ket{l,S-l}$ into the BS, the probability of measuring $k$ and $S-k$ photons behind is the absolute square of a fractional QKT of the input probability amplitudes, $\lvert \mathrm{X}_k \rvert^2= \lvert \sum_{l=0}^S \mathcal{A}^{(r)}_S(k,l) \cdot x_l \rvert^2$, cf.\ Eq.~\ref{eq:KT}. The reflectivity $r$ determines the QKT fractionality, $\alpha=\tfrac{2\theta}{\pi}=\tfrac{4}{\pi}\arcsin\sqrt{r}$. Since two-mode optical interference can be achieved in a single step, regardless of the number of photons involved, this process implements a constant time QKT. For full derivations see SM.
	
	A deeper understanding of the result may be gained from the Schwinger representation of the spin algebra (SM) which links multiphoton interference to spin systems and allows the quantum states to be visualized on a Bloch sphere. In this picture, a total of $S$ photons corresponds to a spin-$\tfrac{S}{2}$ system. The Hamiltonian generating $U_{BS}=\exp\{-i \theta H_{BS}\}$ corresponds to an $S_{x}$ operator for a spin-$\tfrac{S}{2}$. The two-mode Fock state $\ket{l, S-l}$ corresponds an $S_z =\tfrac{S}{2}-l$ eigenstate, known as a Dicke state. Hence, HOM interference may be considered a rotation $R_{\theta,\varphi} = \exp \{-i \theta S_{x} \}$ of $S_z$ around the $S_{x}$ axis on the sphere. It transfers the eigenstate $\ket{\tfrac{S}{2};\tfrac{S}{2}-l}$ to a superposition of Dicke states, Figs.~\ref{Bloch}(\textbf{A}--\textbf{D}). The Q-function in Figs.~\ref{Bloch}(\textbf{E}--\textbf{H}) shows that the initial and final states are eigenstates of two complementary observables, $S_z$ and $S_y$, respectively. Thus, one may identify the former with a position, while the latter with a momentum eigenstate.

\begin{figure}[tbh]\centering
	\includegraphics[height=8.5cm]{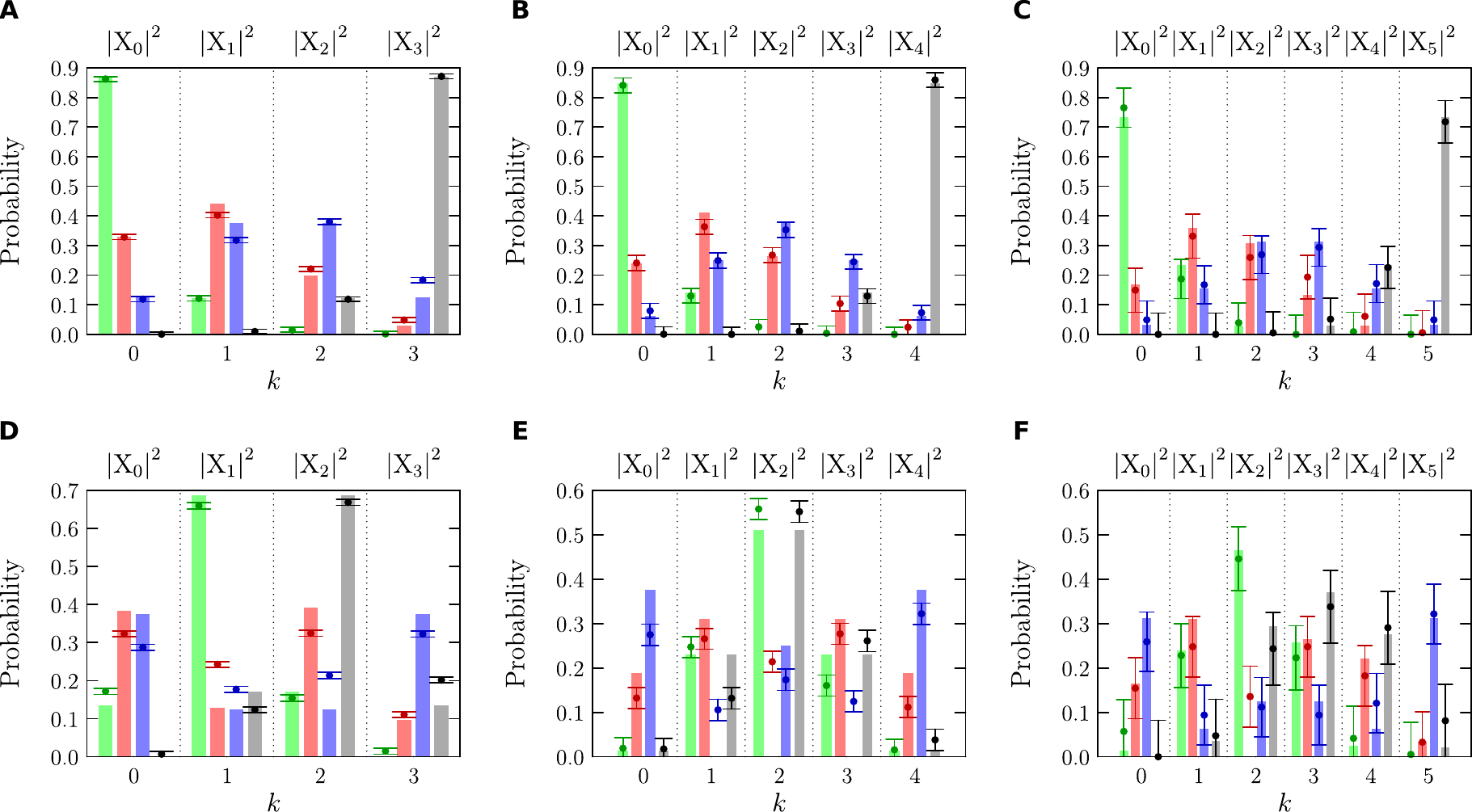}
	\caption{\textbf{Photon number statistics resulting from  Fock state $\ket{l,S-l}$ interference.} The probabilities of detecting $\ket{k}$ and $\ket{S-k}$ photons behind the BS for input (\textbf{A}) $\lvert 0, 3\rangle$, (\textbf{B}) $\lvert 0, 4\rangle$, (\textbf{C}) $\lvert 0,5\rangle$, (\textbf{D}) $\lvert 1, 2\rangle$, (\textbf{E}) $\lvert 2, 2 \rangle$, (\textbf{F}) $\lvert 2,3\rangle$. The BS reflectivities are $r=0.05$ (green), $0.2$ (red), $0.5$ (blue) and $0.95$ (gray). Vertical bars represent theoretical values for an ideal system, while dots are values determined in experiment. The states (\textbf{A}--\textbf{C}) encode sequences $(x_0=1, x_1=0, \dots, x_S=0)$, and in (\textbf{D}) -- $(0, 1, 0, 0)$, (\textbf{E}) -- $(0, 0, 1, 0, 0)$, (\textbf{F}) -- $(0, 0, 1, 0, 0, 0)$, respectively.	The measured probabilities set their QKTs $(\lvert \mathrm{X}_0 \rvert^2, \lvert \mathrm{X}_1 \rvert^2, \dots, \lvert \mathrm{X}_S \rvert^2)$, $\lvert X_k\rvert^2=\lvert \sum_{l=0}^S \mathcal{A}^{(r)}_S(k,l) \cdot x_l \rvert^2$ of fractionality $\alpha = 0.28$ (green), $0.60$ (red), $1.00$ (blue) and $1.72$ (gray).}
	\label{interference}
\end{figure}
	
	The experimental setup for multiphoton HOM interference is depicted in Fig.~\ref{BS}(\textbf{B}). Two pulsed spontaneous parametric down-conversion (SPDC) sources each generate two-mode photon-number correlated states (SM). The signal and idler are separated with a polarization BS (PBS) into four spatial modes. The modes $A$ and $D$ are used for heralding and creation of Fock states $\lvert l \rangle$ in $B$ and $\lvert S-l \rangle$ in $C$ which interfere in a variable ratio fiber coupler (the BS). An optical path delay $\tau$ in one of the pump beams ensures optimal temporal overlap at interference. Photon-number-resolved measurements are achieved using transition edge sensors (TESs) that we previously estimated to achieve over $90\%$ efficiency\cite{Humphreys2015}.
	
	We interfered the vacuum $\lvert 0 \rangle$ ($l=0$) with multiphoton Fock states $\lvert S \rangle$ ($S-l=S$) on a coupler with splitting ratios $r=0.05$ (green), $0.2$ (red), $0.5$ (blue) and $0.95$ (gray), and measured photon number statistics. They are depicted in Figs.~\ref{interference}(\textbf{A}--\textbf{C}) for $S=3,4,5$. The input states encode sequences $(x_0=1, x_1=0, \dots, x_S=0)$, while the measured probabilities set their QKTs: $(\lvert \mathrm{X}_0 \rvert^2, \lvert \mathrm{X}_1 \rvert^2, \dots, \lvert \mathrm{X}_S \rvert^2)$, where $\lvert \mathrm{X}_k \rvert^2= \lvert  \mathcal{A}^{(r)}_S(k,0) \rvert^2$. The reflectivities used correspond to fractionalities $\alpha = 0.28, 0.60, 1.00, 1.72$. Errors were estimated as a square root inverse of the number of measurements (SM). The second-order interferometric visibility reached values between $71.4\%$ and $98.6\%$ for $S=5$ (SM).
	
	For the same values of $r$ we measured photon number distribution resulting from interference of $\ket{1,2}$, $\ket{2,2}$ and $\ket{2,3}$. They are shown in Figs.~\ref{interference}(\textbf{D}--\textbf{F}). The inputs encode $(0, 1, 0, 0)$, $(0, 0, 1, 0, 0)$, $(0, 0, 1, 0, 0, 0)$, while $\lvert \mathrm{X}_k \rvert^2= \lvert  \mathcal{A}^{(r)}_3(k,1) \rvert^2$, $\lvert  \mathcal{A}^{(r)}_4(k,2) \rvert^2$ and $\lvert  \mathcal{A}^{(r)}_5(k,2) \rvert^2$, respectively. The visibility was between $54.8\%$ and $99.5\%$ ($S=5$) (SM).
	
	Fig.~\ref{interference} shows that the theoretical values  computed for an ideal system (the bars) match the experimental results (the dots) well.
	
\section*{Discussion}
\vskip-3mm
	
	Realization of the fractional QKT with qudit systems opens a new prospect for transformation of large data sequences in $O(1)$ time. This is not possible with the implementations based on waveguides. Both cases are examples of a non-universal quantum computer optimized for one task which is the basis for a variety of important applications\cite{Yap2003}.  The photonic proof of concept is currently limited by the range of input states that can be prepared. However, deterministic creation of an arbitrary superposition of Fock states has been demonstrated for trapped ions and superconducting resonators\cite{Hofheinz2009}. Since a BS sees orthogonal spectral or polarization modes independently, one can extend the transform to higher dimensions\cite{Raymer2009,Kobayashi2016}. We note that the QKT could also be implemented on existing quantum annealing processors\cite{DWavePrivateComm}, which operate on a chain of interacting spin-$\tfrac{1}{2}$ systems (SM), and using HOM interference of fermions with symmetric wavefunction of the interfering degrees of freedom. 
	
	Large-scale realizations of the QKT may use an increasing number of measurements as an error minimization strategy (SM). This is akin to the common approach in classical data processing where the accuracy can be improved by an enhanced precision of numeric data types and number of iterations, without altering the proper algorithm. Errors resulting from losses in the system, e.g. at the BS, are easily corrected by applying a postselection scheme and filtering out the cases when the total number of photons behind the BS is lower than in the input state. 
	
	Interestingly, $O(1)$ computation of the fractional FT for continuous variable systems can be implemented with a shallow system realizing a phase shift operation\cite{Fan2012}. This is a quantum counterpart of the operation of a single focusing lens in classical optics, which produces the fractional FT of the image placed at its focal length\cite{Sejdic2010}. The QKT operates on discrete variables, but when the sampling rate and the sequence length of input data increase, the $\alpha$-fractional QKT tends to the $\alpha$-fractional FT\cite{Atakishiyev1997}. This relation is also reflected by the fact that symmetric Kravchuk functions $\phi_k^{(p=1/2)}$ (eigenfunctions of QKT) tend to Hermite--Gauss polynomials (eigenfunctions of the FT) in this limit.
	
	Our result, along with the fact that qudit-based algorithms exhibit significantly lower number of operations than qubit-based ones\cite{Pavlidis2017}, motivates the further development of highly-controllable quantum harmonic oscillator platforms with implications for quantum signal processing in a whole range of applications. Provided efficient input state preparation and detection of larger Fock states, the $O(1)$ QKT demonstrated here in principle may find practical applications in imaging of unprecedented quality, fostering early diagnostics and neuroscience\cite{Cyranoski2017}. 
	
\section*{Materials and Methods}
\vskip-3mm	
	
	A light pulse from a Ti:Sapphire laser at $775\kern.25em\mathrm{nm}$ (FWHM of $2\kern.25em\mathrm{nm}$; repetition rate of $75\kern.25em\mathrm{kHz}$) pumps collinear type-II phase-matched $8\kern.25em\mathrm{mm}$-long SPDC waveguides written in a periodically poled KTP (PP-KTP) crystal sample. They generate two independent photon-number correlated states -- the two-mode squeezed vacua $\ket \Psi = \sum_{n=0}^{\infty} \lambda_n |n,n\rangle $, where $\lambda_n = \tfrac{\tanh^n g}{\cosh g}$ is a probability amplitude for creation of a pair of $n$ photons and $g$ is the parametric gain. The average photon number in the signal and idler mode equals $\braket{\hat{n}} = \sinh^2 g$. For small $g$, $\cosh g \approx 1$, and thus $\lambda_n^2 \approx \sinh^{2n} g = \braket{\hat{n}}^n$. In the experiment, the average photon number is $\braket{\hat{n}} \approx 0.2$. This value is sufficient to ensure the emission of multiphoton pairs, but at the same time to diminish the interferometric visibility of two-photon events. In both output states, the signal and idler pulses are split with a polarization beam-splitter (PBS) to four spatial modes $A$--$D$. Subsequently, they are filtered by bandpass filters with $3\kern.25em\mathrm{nm}$ FWHM angle-tuned to the central wavelength of their respective spectra, in order to reduce the broadband background typically generated in dielectric nonlinear waveguides\cite{Eckstein2011}. The pump beam is discarded with an edge filter. The modes $A$ and $D$ are used for heralding and conditional creation of Fock states in modes $B$ and $C$ which interfere in a variable ratio PM fiber coupler. The coupling ratio can be set in the range $0$-$100\%$ with an error of $\pm1.5\%$. The heralding signal modes (H-pol.) are centered at $1554\kern.25em\mathrm{nm}$, while the interfering idler modes (V-pol.) are at $1546\kern.25em\mathrm{nm}$. We employ transition-edge sensors (TES) running at $70\kern.25em\mathrm{mK}$ which allow for photon-number resolved measurements in all modes\cite{Gerrits2011}. Their voltage output is captured with an ADC card.
	
	Before demonstrating the HOM interference, we characterized the setup. A high photon number resolution and single-mode input states are pivotal for this experiment. The resolution of TES detectors (the confidence that the detector gives a correct information about the number of photons) was previously confirmed to exceed $95\%$\cite{Humphreys2015}. The depth of the HOM dip of $85.9\% \pm 0.3\%$ for a two-photon interference indicates an effective Schmidt mode number $K=1.16$. For the measured 4-tuples of photon numbers losses were computed by assuming perfect setup components, each followed by a beam splitter with a reflection coefficient introducing the loss. We estimate the total transmission in each mode to be approximately $50\%$. For the details, see SM.
	
	Measurements for individual settings of the splitting ratio were taken over approximately $400$ seconds, giving $10^9$ data samples for each $r$ ranging from $0$ to $1$ with a step of approximately $3\%$. Small error bars for low photon numbers and larger bars for the higher ones result from keeping the pump power fixed and near-single-modeness of the interfering beams.

	\clearpage
	
\textbf{Funding:} M.S. and A.B. were supported by the Foundation for Polish Science ``First Team'' project No.\ POIR.04.04.00-00-220E/16-00 (originally: FIRST TEAM/2016-2/17), The National Science Centre (NCN) grant No.\ 2012/04/M/ST2/00789 and MNiSW Iuventus Plus project No.\ IP~2014~044873. A.E., I.W. and M.S. were supported by the Engineering and Physical Sciences Research Council project No.\ EP/K034480/1. Numerical computations were performed with a Zeus cluster in the ACK ``Cyfronet'' AGH computer center.

\textbf{Author contributions:} M.S. and A.B. developed the theory while A.E., M.M., W.R.C., W.S.K., and I.A.W. were responsible for realization of the experiment. J.J.R., S.W.N., T.G., and A.L. delivered and maintained the TES detection system. A.B and A.E. developed the software and performed numerical computations. A.B. prepared the plots. All the co-authors wrote up the manuscript.

\textbf{Competing interests:} The authors declare that they have no competing interests.

\textbf{Data and materials availability:} All data needed to evaluate the conclusions in the paper are present in the paper and/or the Supplementary Materials. Additional data available from authors upon request.

\textbf{Pending patents:} M. Stobińska, A. Buraczewski, A. Eckstein, I. A. Walmsley, \textit{A method of performing quantum Fourier-Kravchuk transform (QKT) and a device configured to implement said method}, patent application No.\ P.426228 at the Patent Office of the Republic of Poland.
	
\renewcommand{\theequation}{S\arabic{equation}}
\setcounter{equation}{0}
\renewcommand{\thefigure}{S\arabic{figure}}
\setcounter{figure}{0}
\renewcommand{\thetable}{S\arabic{table}}
\setcounter{table}{0}

\clearpage

\begin{center}
	{\Large Supplementary Materials}
\end{center}
	
\section{Mathematical foundations for the Kravchuk transform}

\subsection{Kravchuk matrices and polynomials}

Kravchuk polynomials are related to Kravchuk matrices that are generated by a binomial expressions of the form $(1+x)^{N-j}(1-x)^j$, with positive integer $N$ and $j=0,\ldots,N$. Such an expression expands to a set of $N+1$ polynomials $a_{0,j}+a_{1,j}x+a_{2,j}x^2+\ldots+a_{i,j}x^i+\ldots+a_{N,j}x^N$. Their coefficients can be used to construct a square $(N+1)\times(N+1)$ matrix $\textbf{K}^{(N)}$ with an entry in $i$th row and $j$th column given by $\textbf{K}^{(N)}_{i,j}=a_{i,j}$. Thus,
\begin{equation}
(1+x)^{N-j}(1-x)^j = \sum_{i=0}^{N} x^i\, \textbf{K}^{(N)}_{i,j}.
\end{equation}

As an example, let us examine the expression $(1+x)^{N-j}(1-x)^j$ for $N=3$ and $j=0,\dots,3$. Its all possible expansions read
\begin{align*}
j=0:&& (1+x)^3={}& 1+3x+3x^2+x^3,\\
j=1:&& (1+x)^2(1-x)={}& 1+x-x^2-x^3,\\
j=2:&& (1+x)(1-x)^2={}& 1-x-x^2+x^3,\\
j=3:&& (1-x)^3={}& 1-3x+3x^2-x^3,
\end{align*}
and thus, the corresponding Kravchuk matrix is as follows
\begin{equation*}
\textbf{K}^{(3)} = \begin{pmatrix}
1& 1& 1& 1\\
3& 1& -1& -3\\
3& -1& -1& 3\\
1& -1& 1& -1\\
\end{pmatrix}.
\end{equation*}

Kravchuk matrices possess many interesting properties\cite{Feinsilver2005}. The top row contains only 1's, while the bottom one $(-1)^j$. The first column is expressed by binomial coefficients $\binom{N}{i}$ and the last one by $(-1)^i\binom{N}{i}$. The matrix is also characterized by a four-fold symmetry $\lvert\textbf{K}^{(N)}_{i,j}\rvert=\lvert\textbf{K}^{(N)}_{N-i,j}\rvert=\lvert\textbf{K}^{(N)}_{i,N-j}\rvert=\lvert\textbf{K}^{(N)}_{N-i,N-j}\rvert$. Finally, $\left(\textbf{K}^{(N)}\right)^2=2^N\,\mathbf{I}$, where $\mathbf{I}$ is the identity matrix. The latter property is important as it underpins the existence of the Kravchuk transform. Kravchuk matrices are considered to be generalized Pascal triangles, while the columns of $\textbf{K}^{(N)}$ are called generalized binomial coefficients. The entries of $\textbf{K}^{(N)}$ can be computed using the following formula
\begin{equation}
\textbf{K}^{(N)}_{i,j} = \sum_{k=0}^{i} (-1)^k\binom{j}{k}\binom{N-j}{i-k}=k^{(1/2)}_i(j,N),
\end{equation}
where $k^{(1/2)}_n(x,N)$ denotes $n$-th symmetric Kravchuk polynomial of variable $x$ and order $N$.

Kravchuk polynomials are defined by the rows of $\textbf{K}^{(N)}$. Their domain is $x=0,\ldots,N$. They fulfill the following orthogonality relation
\begin{equation}
\frac{1}{2^N}\sum_{j=0}^N\binom{N}{j} k^{(1/2)}_n(j,N)\, k^{(1/2)}_m(j,N) = \frac{1}{2^{2n}}\binom{N}{n}\delta_{n,m},
\end{equation}
where $n$ and $m$ are integers and $\delta_{m,n}$ is the Kronecker delta equal 1 where $n=m$ and 0 otherwise.

In case of an unsymmetric binomial expression, the Kravchuk matrix can be generated like this
\begin{equation}
(1+(p-1)x)^{N-j}(1-x)^j = \sum_{i=0}^N x^i \mathbf{K}_{i,j}^{(p,N)} = \sum_{i=0}^N x^i k_i^{(p)}(j,N),
\end{equation}
where $0\leq p\leq 1$ and the Kravchuk polynomials 
\begin{equation}
k_i^{(p)}(j,N) = \sum_{k=0}^{i} (-1)^k(p-1)^{i-k}\binom{j}{k}\binom{N-j}{i-k}
\end{equation}
fulfill the following orthogonality relation
\begin{equation}
\sum_{j=0}^N\binom{N}{j}(p-1)^j k^{(p)}_n(j,N)\, k^{(p)}_m(j,N) = p^N(p-1)^n\binom{N}{n}\delta_{n,m}.
\end{equation}
Kravchuk polynomials belong to the family of special functions and can also be expressed in terms of a Gauss hypergeometric function ${}_2F_1[a,b;c;z]$ (see Section \ref{Appendix})
\begin{equation}
k^{(p)}_n(x,N) = (-1)^n\binom{N}{n} p^n\,{}_2F_1\left[-n,-x;-N;\tfrac{1}{p}\right]
= (-1)^n\binom{N}{n} p^n \sum_{i=0}^{n}\binom{n}{i}(-1)^i \dfrac{(-x)_i}{(-N)_i}\,\dfrac{1}{i!\,p^i},
\label{eq:krav2F1}
\end{equation}
where $(a)_i=a\cdot(a+1)\cdot\ldots\cdot(a+i-1)$ is the Pochhammer symbol. From Eq.~\ref{eq:krav2F1} it is clear that $k^{(p)}_n(x,N)$ are polynomials of degree $n$ and that the parameter $n$ and variable $x$ can be exchanged in the following way
\begin{equation}
(-1)^x\binom{N}{x} p^x k^{(p)}_n(x,N) = (-1)^n\binom{N}{n} p^n k^{(p)}_x(n,N).
\end{equation}
Kravchuk polynomials can be regarded as a discrete and finite counterpart of Hermite polynomials $H_n(x)$\cite{Atakishiyev1997}
\begin{equation}
\lim_{N\to\infty}\left(\dfrac{2}{Np(1-p)}\right)^{n/2}
n!\,k_n^{(p)}\left(Np+x\sqrt{2Np(1-p)},N\right)=H_n(x).
\end{equation}

\begin{figure}\centering
	\includegraphics[width=13cm]{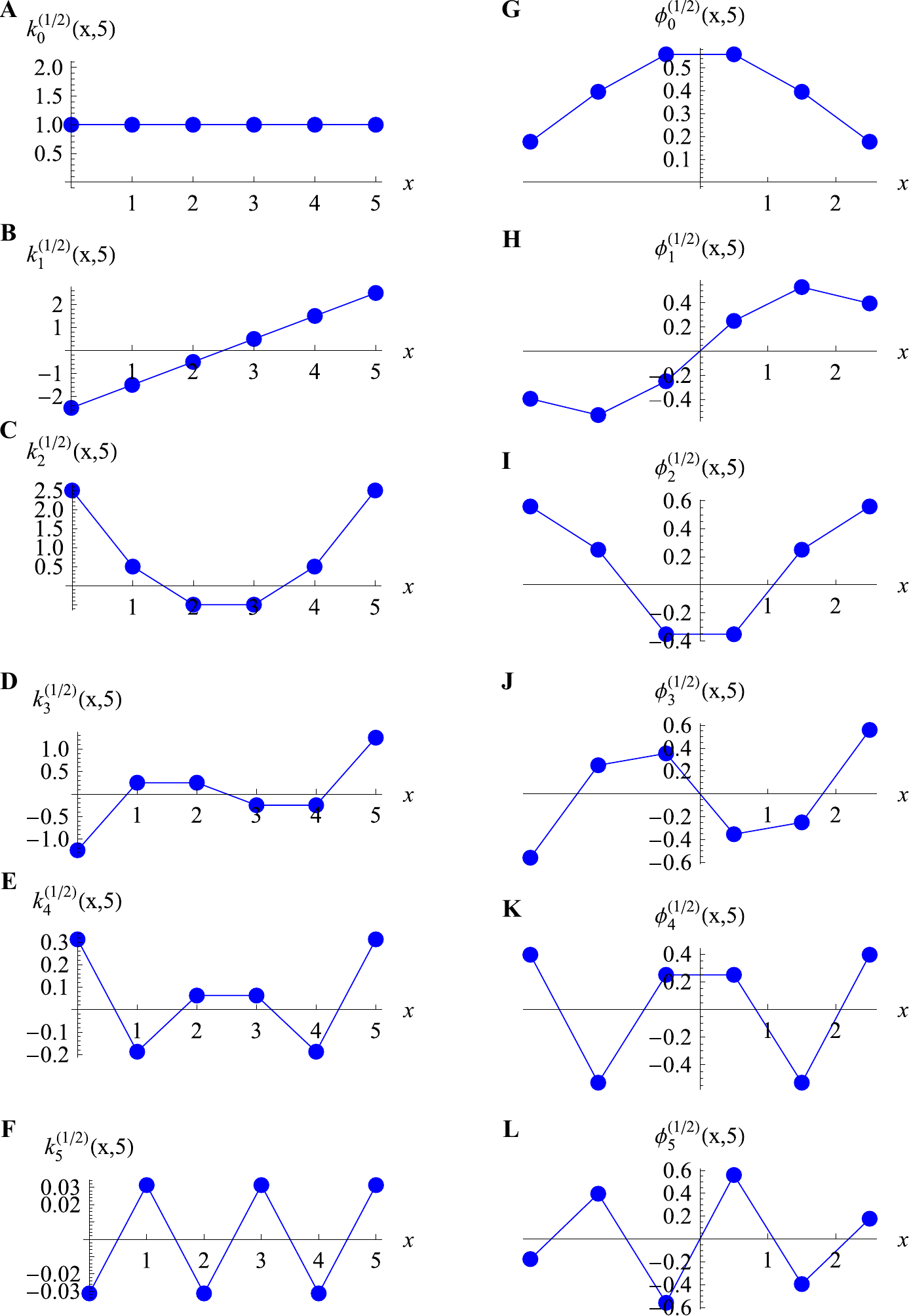}
	\caption{\textbf{Symmetric Kravchuk polynomials $k_n^{(1/2)}(x,N)$ and functions $\phi_n^{(1/2)}(x,N)$}. The plots  (\textbf{A})-(\textbf{F}) show Kravchuk polynomials, while (\textbf{G})-(\textbf{L}) -- Kravchuk functions. Computations are performed for $N=5$.}
	\label{fig:kravpoly}
\end{figure}

\subsection{Kravchuk functions}

Kravchuk functions $\phi_n^{(p)}$ are Kravchuk polynomials that are normalized and centered at the maximum of a binomial distribution $\binom{N}{x}\,p^{x} (1-p)^{N-x}$
\begin{equation}
\phi_n^{(p)}(x-Np,N) = \sqrt{\dfrac{n!(N-n)!}{x!(N-x)!}}\sqrt{p^{x-n}(1-p)^{N-n-x}}k_n^{(p)}(x,N).
\end{equation}
Their domain is $x=-Np,-Np+1,\ldots,-Np+N$.  They are orthonormal
\begin{equation}
\sum_{i=0}^N \phi_n^{(p)}(i-Np,N)\,\phi_m^{(p)}(i-Np,N)=\delta_{n,m}
\end{equation}
and the variables $n$ and $i$ can be used interchangeably
\begin{equation}
\phi_n^{(p)}(i-Np,N) = \phi_i^{(p)}(n-Np,N).
\end{equation}
Interestingly, Kravchuk functions are solutions of finite oscillator wave equation
\begin{equation}
\mathbf{H}^{(N)}(x)\,\phi_n^{(1/2)}(x,N) = (n+\tfrac{1}{2})\,\phi_n^{(1/2)}(x,N),\qquad n=0,\ldots,N,
\end{equation}
where $\mathbf{H}^{(N)}$ is a finite-difference operator identified as a discrete Hamiltonian of this system. In the limit of $N\to\infty$ Kravchuk functions tend to the harmonic oscillator wave functions $\psi_n(x)$ (Hermite--Gauss polynomials)
\begin{equation}
\lim_{N\to\infty} (N/2)^{1/4} \phi_n^{(1/2)}\left(x\sqrt{N/2},N\right)=\psi_n(x),
\end{equation}
where
\begin{equation}
\psi_n(x)=\dfrac{1}{\sqrt{\sqrt{\pi}\,2^n\,n!}}\,H_n(x)\,e^{-x^2/2}.
\end{equation}
Fig.~\ref{fig:kravpoly} shows Kravchuk polynomials and Kravchuk functions for $N=5$.

\subsection{Kravchuk transform}

The Kravchuk transform (KT) is defined by means of Kravchuk functions. KT converts an input sequence $(x_0,x_1,\dots,x_S)$ into a new string $(X_0,X_1,\dots,X_S)$ in the following way (cf.\ the main text)
\begin{equation}
X_k=\sum_{l=0}^S e^{-i\frac{\pi\alpha}{2}\,\tfrac{S}{2}}
e^{i\frac{\pi}{2}(l-k)}\,\phi_k^{(p)}(l-Sp,S)\cdot x_l,\qquad k=0,\ldots,S,
\label{eq:KT_sup}
\end{equation}
where $\alpha$ is the fractionality of the transform and $p=\sin^2(\pi\alpha/4)$. Effectively, it decomposes the input string in the basis of Kravchuk functions. The KT can also be seen as multiplication of the input vector by a rescaled Kravchuk matrix $\mathbf{K}^{(N)}$ with additional phase terms.

\begin{figure}\centering
	\includegraphics[width=15cm]{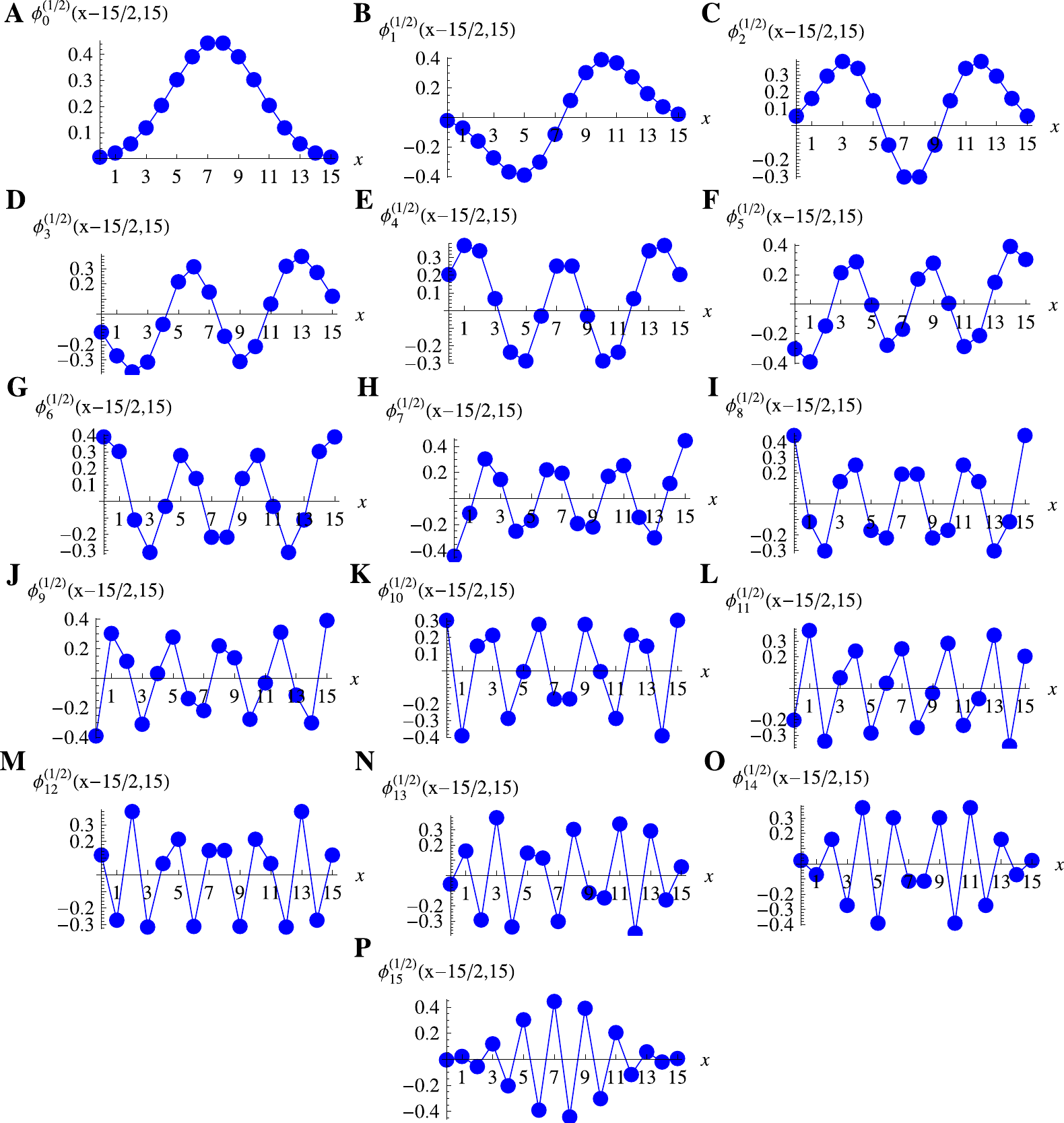}
	\caption{\textbf{Basis states for a 16-point Kravchuk transform}. Plots (\textbf{A})-(\textbf{P}) depict orthonormal basis states $\phi_k^{(1/2)}(x-S/2,S)$ for $S=15$ and $k=0,\ldots,S$, respectively.}
	\label{fig:kravbasis}
\end{figure}

As an illustration, Fig.~\ref{fig:kravbasis} presents the full set of basis states of a 16-point Kravchuk transform ($S=15$ and $k=0,\ldots,15$). 

\clearpage

The parameter $\alpha$ modifies the Kravchuk transform (Eq.~\ref{eq:KT_sup}) in the following way
\begin{enumerate}
	\item for $\alpha=1$, it is a forward transform,
	\item for $\alpha=2$, it is an inverse (backward) transform, allowing one to compute the original sequence $(x_l)$ from $(X_k)$ using the same algorithm or a physical system, 
	\item for $\alpha=3$, the transform negates the input, $X_k=-x_k$,
	\item for $\alpha=0$ or for $\alpha=4$, the transform is an identity operation, $X_k=x_k$.
\end{enumerate}
Similar properties are observed for the (integral) Fourier transform. 

In addition, the Fourier transform and the KT can be seen as circular rotations of the data in the time-frequency space by an angle $\theta=\pi\alpha/2$. This rotation is also well-defined for $0<\alpha<1$, leading to an $\alpha$-fractional transforms. The KT is additive with respect to $\alpha$, i.e.\ application of two consecutive transforms parameterized by $\alpha_1$ and $\alpha_2$ results in a $\alpha$-fractional KT with $\alpha=\alpha_1+\alpha_2$.

\subsection{Discrete Fourier transform}

\begin{figure}\centering
	\includegraphics[width=15cm]{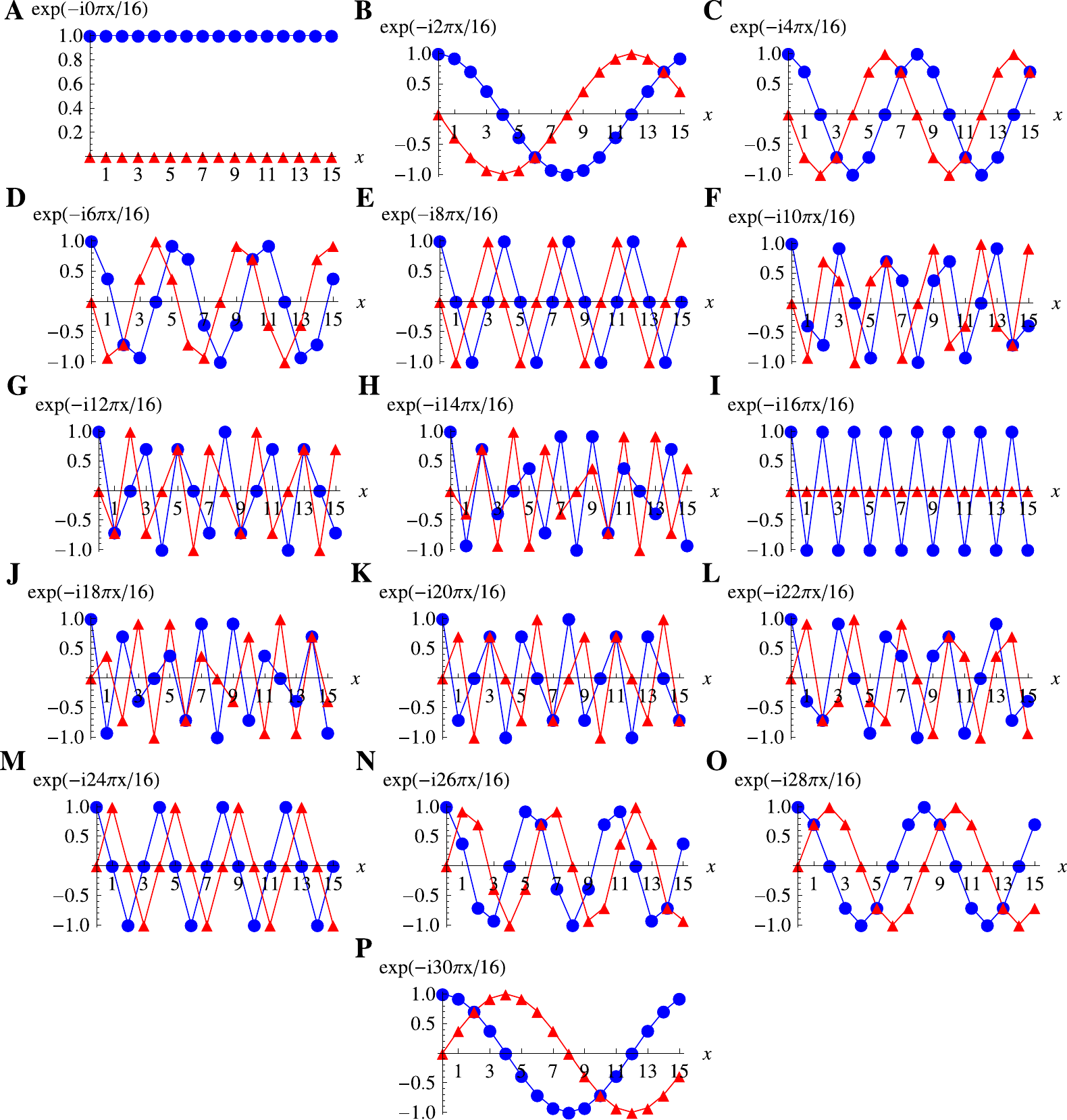}
	\caption{\textbf{Basis states for a 16-point discrete Fourier transform}. Plots (\textbf{A})-(\textbf{P}) depict 16 basis states $\exp\{-i2\pi k / (S+1)\}$ for $S=15$ and $k=0,\ldots,S$. Blue circles denote real, while red triangles -- imaginary components of the states.}
	\label{fig:DFTbasis}
\end{figure}

The discrete Fourier transform (DFT) as well as the Fast Fourier Transform (FFT) algorithm decompose input data string in the basis of plane waves
\begin{equation}
X_k = \frac{1}{\sqrt{S+1}} \sum_{l=0}^S e^{-i2\pi\frac{kl}{S+1}} \cdot x_l.
\label{eq:DFT_sup}
\end{equation}
Fig.~\ref{fig:DFTbasis} depicts the full set of basis states of a 16-point discrete Fourier transform ($S=15$ and $k=0,\ldots,15$). 

\begin{figure}\centering
	\includegraphics[width=15cm]{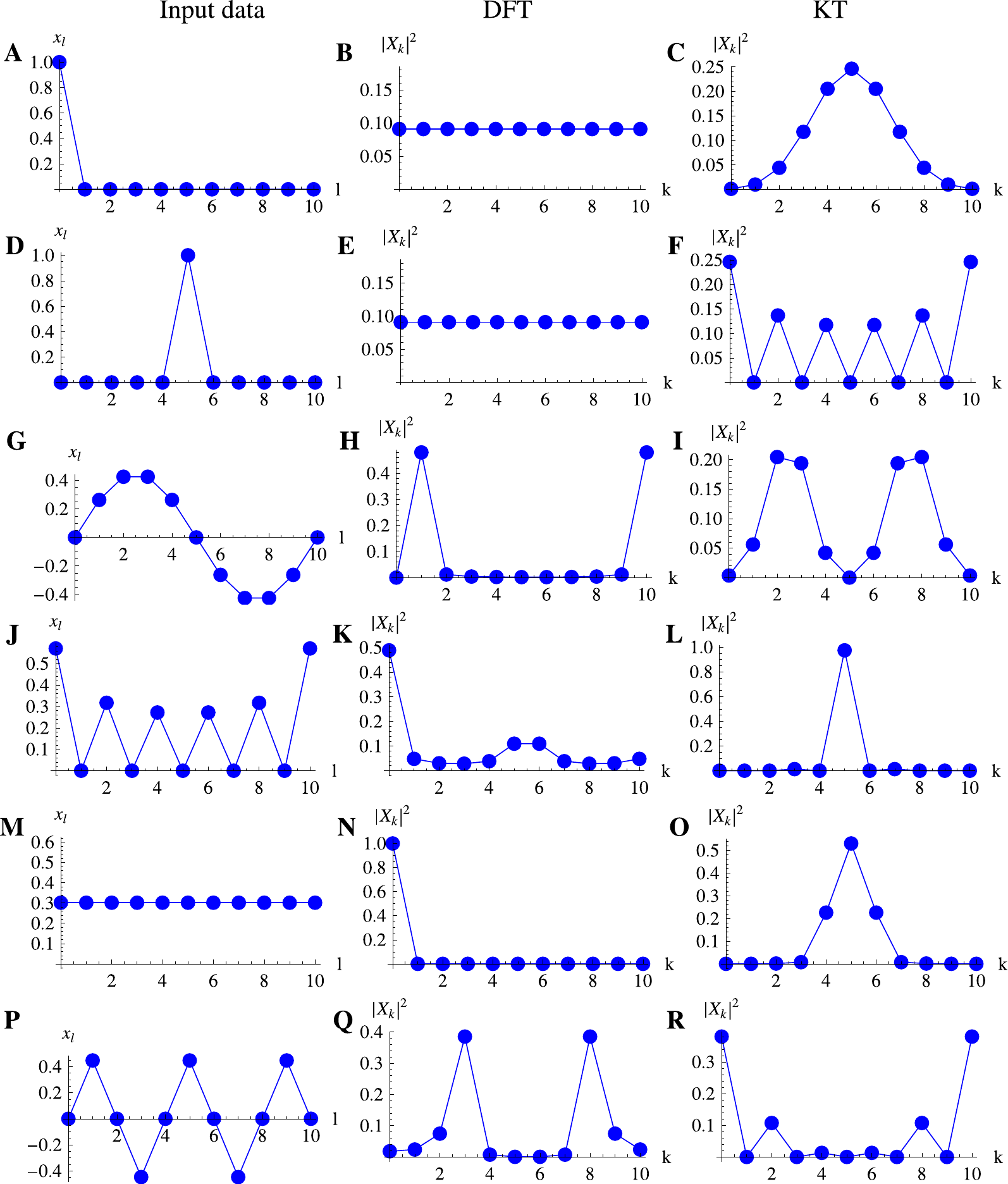}
	\caption{\textbf{Kravchuk vs. discrete Fourier transform.} The left column depicts exemplary input data $(x_l)$ for $l=0,\ldots,10$, while the middle and right columns -- the results of computation of $\lvert X_k\rvert^2$ using the DFT and the Kravchuk transform, respectively.}
	\label{fig:KTvsFFT}
\end{figure}

Since plane waves $e^{-i2\pi\frac{kl}{S+1}}$ are periodic and their domain consist of negative and positive integers and zero, the discrete Fourier transform defined in Eq.~\ref{eq:DFT_sup} correctly decomposes only samples of periodic data. Since the Fourier transform of a discrete periodic signal with period of length $S+1$ results also in a discrete periodic output of the same length, it is enough to process one period of the input string, $(x_0,\ldots,x_S )$, and store one period of the output sequence $(X_0,\ldots\,X_S)$. However, one should remember that effectively the input sequence is seen by this algorithm as infinite
\begin{displaymath}
\ldots,x_0,x_1,\ldots,x_S,x_0,x_1,\ldots,x_S,x_0,x_1,\ldots,x_S,\ldots
\end{displaymath}
and so is the resulting sequence
\begin{displaymath}
\ldots,X_0,X_1,\ldots,X_S,X_0,X_1,\ldots,X_S,X_0,X_1,\ldots,X_S,\ldots
\end{displaymath}
As a result of this, two shifted input sequences, e.g.\ $(1,0,0,0,0,0,0,0,0)$ (Fig.~\ref{fig:KTvsFFT}(\textbf{A})) and $(0,0,0,0,1,0,0,0,0)$ (Fig.~\ref{fig:KTvsFFT}(\textbf{D})), are transformed by the DFT to the output differing only by a complex phase (Fig.~\ref{fig:KTvsFFT}(\textbf{B}) \& (\textbf{E})). Figs.~\ref{fig:KTvsFFT} shows DFT and KT for exemplary input sequences.

$\alpha$-fractional DFT (also known as DFRFT) for $0<\alpha\leq 1$ is defined as\cite{Sejdic2010}
\begin{equation}
X_k = \sqrt{\frac{\sin\theta-i\cos\theta}{S+1}}
e^{\frac{i}{2}k^2\cot\theta}
\sum_{l=0}^S e^{-i2\pi\frac{kl}{S+1}}
e^{\frac{i}{2}l^2\cot\theta}
\cdot x_l,
\label{eq:DFRFT}
\end{equation}
where $\theta=\frac{\pi\alpha}{2}$.  For $\alpha=1$ Eq.~\ref{eq:DFRFT} reduces to Eq.~\ref{eq:DFT_sup}.

\subsection{KT and DFT software implementations}

Algorithms for computation of Kravchuk polynomials and transforms as well as the FFT have been proposed both for software and hardware solutions. The most known library for computation of the KT is POLPAK, while for the FFT it is FFTW. KT algorithms underperform the FFT in speed because their number of operations is $O(n^2\log^2 n)$ compared to $O(n\log n)$ for the FFT. This result has been improved by Venkataramana et al. to $O(n^2)$ by using the Clenshaw's recurrence formula\cite{Venkataramana2011}. Nevertheless, this seriously limits application of the KT in data processing. For example, an image of $512\times 512$ pixels is transformed with the KT in $15\kern.25em\mathrm{min}$ instead of several seconds as is in case of the use of FFT\cite{Papakostas2011}. 

\section{Applications of Kravchuk transform}
\label{sec:applications}

Following the seminal paper by Yap et al.\cite{Yap2003}, numerous applications of the Kravchuk functions, polynomials and the KT have been proposed.

\subsection{Image processing}

In computer image processing, image moments are vectors or matrices which describe interesting properties of the source data. They can be obtained e.g.\ by decomposition of pixel intensities in the basis of orthogonal functions. Kravchuk polynomials have been found to be extremely useful for this purpose, as they are well defined on a finite domain of raster images and the computed moments carry out a lot of information about characteristic image features. In practice, computing Kravchuk moments is equivalent to performing the Kravchuk transform of the input image. This method has been already shown to be useful in optical character recognition\cite{Yap2003}, autonomous reading of the sign language, hand signature discrimination, writer identification, automatic face and gesture recognition\cite{Gautam2017} as well as facial expression and gait analysis. Importantly, this approach allows one to extract both local and global features of the input data and to recognize images corrupted with noise or tilt and possessing change in e.g.\ facial expression\cite{Gautam2017}.

The $\alpha$-fractional KT has been shown to be very useful in data watermarking schemes (watermark insertion and detection)\cite{Liu2017}. It allows to produce results which are invariant to image rotation, scaling and translation. For these reasons, the KT can be used in e.g.\ detection of copy-move forgery of images.

\subsection{Search engines}

The vectors obtained by the decomposition of the data in the basis of the Kravchuk functions have been also successfully tested in search engines. Such descriptors are especially able to capture sharp changes in the source data. By using only low-order Kravchuk moments one is able to perform e.g.\ search for similarly-looking 2- and 3-dimensional shapes\cite{Mademlis2006}, detect types of objects seen in a radar signal, perform automatic classification of videos, still images (including medical images) and sound files. Kravchuk transform has been also proposed as a method of efficient lossy compression of images, similarly to the discrete cosine transform used e.g.\ in JPEG file format.

\subsection{Medical image recognition}

In biology, Kravchuk image moments have been tested for determination of drugs in human plasma microscopic images and prediction of  phosphorylation sites in cells. This approach has been thoroughly compared with the state-of-the-art methods towards various medical imaging applications\cite{Papakostas2009}. It was chosen as the best performing for breast mammography images, where it allowed to identify benign and malign masses with 90\% accuracy, compared to 81\% offered by the other techniques. This scheme was proven to outperform also other algorithms in analysis of computer tomography and ultrasound scans towards recognition of liver and prostate tumors.

\subsection{Medical image reconstruction}

KT has been already tested as a potential replacement of the FFT in generation of diagnostic ultrasound and magnetic resonance images (MRI). In these systems, the data regarding patients' bodies are collected in the frequency domain \hbox{(k-space)} and next, are processed by the inverse Fourier transform to obtain a real-space image. Tests were performed with MRI data coming from open repositories of brain and knee examinations. The images were reconstructed with the Kravchuk, Zernike, Pseudo-Zernike, Fourier-Merlin, Legendre and Chebyshev kernels\cite{Yap2003,Papakostas2009}. It has been pointed out that only the Kravchuk and Chebyshev transforms are discrete and allow to operate in the original Cartesian image coordinates. The Kravchuk-based method presented the best behavior in most test cases, giving the smallest reconstruction error and the highest peak signal-to-noise ratio as the moment order increased thus, is best suited for processing of high resolution data. 

\begin{figure}[htp]\centering
	\includegraphics[width=15cm]{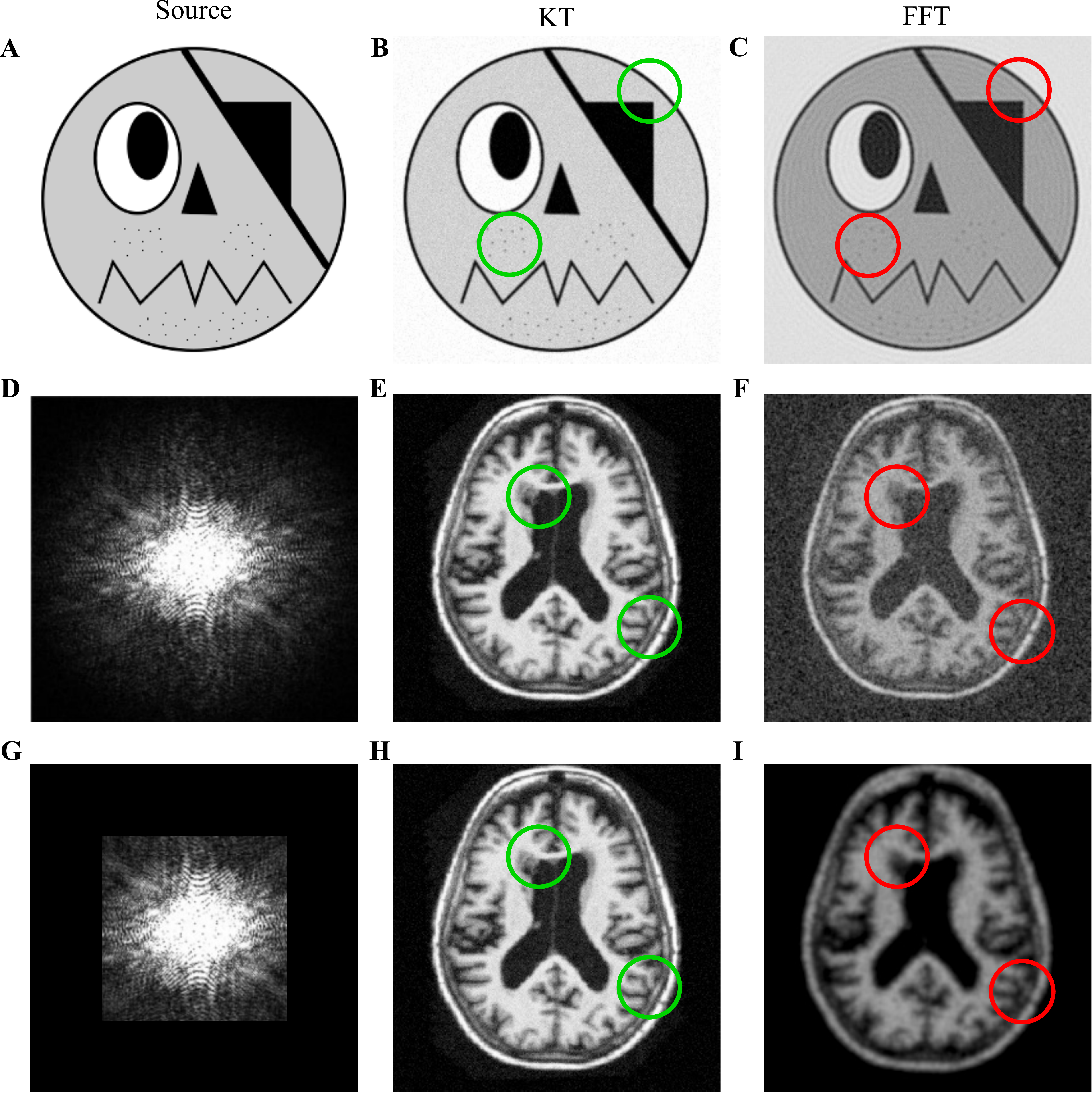}
	\caption{\textbf{Example of FFT and KT image processing}. Two $512\times 512$-pixel test images, a ``pirate'' (\textbf{A}) and a brain scan, the latter in a form of a raw k-space data from the OASIS database (\textbf{D}), were used. First, the ``pirate'' image has been transformed into the k-space with the FFT algorithm to keep both inputs in the same form. Additionally, a third set of data was created by truncating the brain raw data to $256\times 256$ values by removing higher-frequency components (\textbf{G}). Next, all the images in the k-space were supplemented with a $1\%$ additive white Gaussian noise, and reconstructed with corresponding inverse transforms to model the operation of an MRI analysis. (\textbf{B}), (\textbf{E}) \& (\textbf{H}) are the images reconstructed with the KT. The green circles mark some fine details which were retained during this processing. (\textbf{C}), (\textbf{F}) \& (\textbf{I}) are the images reconstructed using the FFT. The red circles highlight the artifacts.}
	\label{brains}
\end{figure}

To assess the advantage of the KT for the MRI diagnostics, we performed similar steps to the ones presented in\cite{Papakostas2009}. Fig.~\ref{brains} shows a comparative numerical study of the KT and FFT for a ``pirate'' test image and a brain scan from the OASIS database. Both source figures have resolution of $512\times 512$ pixels. The source image of the brain is made of k-space raw data from an MRI system, while the ``pirate'' was originally prepared in the real space and next converted to the k-space using the NumPy numerical library to keep both data sets in the same form. A third data set consisted of the original brain scan but truncated to $256\times 256$ values in the k-space by removing (zeroing) the higher-frequency components.

Subsequently, a $1\%$ white Gaussian noise was added in parallel to all three data sets (still in the k-space). No further distortion was applied to any of the figures. Finally, the k-space images were transformed to the real space (reconstructed) with the FFT and KT.

In our test, the FFT produced artifacts, while some details (which could be tumor cells) were missing. This is best captured by the structural similarity index (SSIM). For the ``pirate'' it was $0.73$ (FFT) and $0.92$ (KT), and for the $512\times 512$ brain it was $0.84$ and $0.98$, respectively. In case of the $256\times 256$ brain figure, SSIM achieved was $0.72$  for the FFT and $0.91$ for the KT. The mean square error (MSE) was ten times smaller and the peak signal-to-noise ratio (PSNR) was $10\kern.25em\mathrm{dB}$ larger for the KT than for the FFT. The FFT led to degradation of the usable resolution from $1$-$2\kern.25em\mathrm{mm}$ per voxel to over $5\kern.25em\mathrm{mm}$. Our findings confirmed the results of the previous research\cite{Papakostas2009}. 

\section{Theory: multi-photon Hong--Ou--Mandel interference}

We will now analyze in a detailed manner a generalized multi-photon HOM effect. As explained in the main text, we consider two interfering modes $a$ and $b$ on a beam splitter device with a tunable reflectivity $r$ (defined as the probability of reflection of a single photon). As the input states we take photon number (Fock) states $\ket{l}_a \!\!=\!\! \tfrac{(a^{\dagger})^{l}}{\sqrt{l!}} \ket{0}_a$ and $\ket{S-l}_b\!\!=\!\! \tfrac{(b^{\dagger})^{S-l}}{\sqrt{(S-l)!}} \ket{0}_b$.

\subsection{The Schwinger representation} 

One may represent the su(2) Lie algebra in terms of the annihilation and creation operators of the harmonic oscillator -- the Schwinger representation. For a single spin two independent oscillators $a$ and $b$ are required. The spin operators are then constructed in the following way
\begin{equation}
S_x=\dfrac{a^\dagger b+a\,b^\dagger}{2},\qquad
S_y=\dfrac{i\left(a\,b^\dagger-a^\dagger b\right)}{2},\qquad
S_z=\dfrac{a^\dagger a-b^\dagger b}{2},\qquad
S_0=\dfrac{a^\dagger a+b^\dagger b}{2}.
\end{equation}
$S_0$ is the Casimir operator $S_0 (S_0+1) = S_x^2+S_y^2+S_z^2$. The spin components fulfill the standard su(2) commutation relations
\begin{equation}
[S_x, S_y] = iS_z, \qquad [S_y,S_z] = iS_x, \qquad [S_z,S_x] = iS_y.
\end{equation}

\subsection{Beam splitter}

Interference of two independent modes $a$ and $b$ on a beam splitter is governed by the following Hamiltonian
\begin{align}
H={}& H_0+H_{BS},\\
H_0={}& \tfrac{\hbar}{2}\left(a^\dagger a+b^\dagger b\right),\\
H_{BS}={}&\tfrac{i\hbar}{2}\left(a^\dagger b\,e^{-i\varphi}-a\,b^\dagger e^{i\varphi}\right).
\end{align}
$H_0$ is the free quantum oscillator energy and $H_{BS}$ -- the beam splitter interaction\cite{Kim2002}. $\varphi$ is the phase difference between the reflected and transmitted fields behind the beam splitter. $H_0$ commutes with $H_{BS}$.

\noindent
Using the Schwinger representation, we express $H$ in terms of the spin operators $S_0, S_x, S_y, S_z$
\begin{align}
H_0 = \hbar S_0,
\end{align}
\begin{align}
a^\dagger b\,e^{-i\varphi}-a\,b^\dagger e^{i\varphi}
{}={}&
\cos\varphi\left(a^\dagger b\,-a\,b^\dagger\right)-i\sin\varphi\left(a^\dagger b\,+a\,b^\dagger\right)
\\
{}={}&
2i\left(\cos\varphi\cdot S_y-\sin\varphi\cdot S_x\right).
\end{align}
\begin{align}
H_{BS}={}& \tfrac{i\hbar}{2}\,2i\left(\cos\varphi\cdot S_y-\sin\varphi\cdot S_x\right)
\\
{}={}&
\hbar\left(\sin\varphi\cdot S_x-\cos\varphi\cdot S_y\right).
\end{align}
The Hamiltonian generates the evolution operator
\begin{align}
U={}&\exp\{-i\theta H/\hbar\}
\\
{}={}& \exp\{-i\theta(H_0+H_{BS})/\hbar\}
\\
{}={}& \exp\{-i\theta\,H_{BS}/\hbar\} \exp\{-i\theta\,H_0/\hbar\}
\\
{}={}& U_{BS}\,U_0,
\\
U_0={}&\exp\left\{-i\theta S_0 \right\},
\\
U_{BS}={}&\exp\left\{-i\theta\left(\sin\varphi\cdot S_x-\cos\varphi\cdot S_y\right)\right\}.
\end{align}

\noindent
The evolution in the Heisenberg picture allows to establish a linear relation between the input $(a,b)$ and the output $(a_r, a_t)$ annihilation operators
\begin{align}
a_r = U_{BS}^{\dagger}\, a\, U_{BS} = a \cos\tfrac{\theta}{2} + b \,e^{-i\varphi} \sin\tfrac{\theta}{2}, \\
a_t = U_{BS}^{\dagger}\, b\,  U_{BS} = - a \,e^{i\varphi} \sin\tfrac{\theta}{2} + b \cos\tfrac{\theta}{2}.
\end{align}
The relation takes the following matrix form
\begin{align}
\mathbf{U_{BS}}={}&\begin{pmatrix}
\cos\tfrac{\theta}{2} & \sin\tfrac{\theta}{2} \,e^{-i\varphi}\\
-\sin\tfrac{\theta}{2} \,e^{i\varphi}& \cos\tfrac{\theta}{2} 
\end{pmatrix},
\end{align}
where $\mathbf{U_{BS}}\,\mathbf{U_{BS}^\dagger}=\mathbf{1}$ and $\mathbf{U_{BS}^\dagger}=\mathbf{U_{BS}}^{-1}$ hold true. $U_0$ amounts to a global phase. 

We now substitute $\sin\tfrac{\theta}{2} =\sqrt{r}$ and $\cos\tfrac{\theta}{2} =\sqrt{1-r}$ to relate the evolution directly to the beam splitter reflectivity
\begin{align}
\mathbf{U_{BS}}={}&\begin{pmatrix}
\sqrt{1-r}& e^{-i\varphi}\sqrt{r}\\
-e^{i\varphi}\sqrt{r}& \sqrt{1-r}
\end{pmatrix},
\end{align}
\begin{align}
\mathbf{U_{BS}}^{-1}=\mathbf{U_{BS}^\dagger}={}&\begin{pmatrix}
\sqrt{1-r}& -e^{-i\varphi}\sqrt{r}\\
e^{i\varphi}\sqrt{r}& \sqrt{1-r}
\end{pmatrix}.
\end{align}
This brings us to the following relation between the input and output creation operators, to be used in the next section
\begin{align}
\begin{pmatrix}a\\b\end{pmatrix}={}&
\mathbf{U_{BS}}^{-1}\begin{pmatrix}a_r\\a_t\end{pmatrix}
=
\begin{pmatrix}
\sqrt{1-r}& -e^{-i\varphi}\sqrt{r}\\
e^{i\varphi}\sqrt{r}& \sqrt{1-r}
\end{pmatrix}
\begin{pmatrix}a_r\\a_t\end{pmatrix},
\end{align}

\begin{align}
\begin{pmatrix}a^\dagger\\b^\dagger\end{pmatrix}={}&
\begin{pmatrix}
\sqrt{1-r}& -e^{i\varphi}\sqrt{r}\\
e^{-i\varphi}\sqrt{r}& \sqrt{1-r}
\end{pmatrix}
\begin{pmatrix}a_r^\dagger\\a_t^\dagger\end{pmatrix},
\end{align}

\begin{align}
a^\dagger={}& \sqrt{1-r}\,a_r^\dagger-e^{i\varphi}\sqrt{r}\,a_t^\dagger,\\
b^\dagger={}& e^{-i\varphi}\sqrt{r}\,a_r^\dagger+\sqrt{1-r}\,a_t^\dagger.
\end{align}

\subsection{Photon number amplitude}

Let the input states in modes $a$ and $b$ be the Fock states $\ket{l}$ and $\ket{S-l}$, respectively. Then,
\begin{align}
U_0\ket{l}_a\ket{S-l}_b={}& e^{-i\theta \tfrac{S}{2}}\ket{l}_a\ket{S-l}_b,
\\
U_{BS}\ket{l}_a\ket{S-l}_b={}&
U_{BS}
\dfrac{\left(a^\dagger\right)^l}{\sqrt{l!}}
\dfrac{\left(b^\dagger\right)^{S-l}}{\sqrt{(S-l)!}}
\ket{0}
\\
{}={}&
\dfrac{1}{\sqrt{l!\,(S-l)!}}
\left(\sqrt{1-r}\,a_r^\dagger-e^{i\varphi}\sqrt{r}\,a_t^\dagger\right)^l
\left(e^{-i\varphi}\sqrt{r}\,a_r^\dagger+\sqrt{1-r}\,a_t^\dagger\right)^{S-l}
\ket{0}
\\
{}={}&
\dfrac{1}{\sqrt{l!\,(S-l)!}}
\begin{aligned}[t]
&\sum_{m=0}^l
\sum_{n=0}^{S-l}
\binom{l}{m}
\binom{S-l}{n}
\left(\sqrt{1-r}\,a_r^\dagger\right)^m
\left(-e^{i\varphi}\sqrt{r}\,a_t^\dagger\right)^{l-m}
\times{}\\
&{}\times
\left(e^{-i\varphi}\sqrt{r}\,a_r^\dagger\right)^n
\left(\sqrt{1-r}\,a_t^\dagger\right)^{S-l-n}
\ket{0}
\end{aligned}
\\
{}={}&
\dfrac{1}{\sqrt{l!\,(S-l)!}}
\begin{aligned}[t]
&\sum_{m=0}^l
\sum_{n=0}^{S-l}
\binom{l}{m}
\binom{S-l}{n}
\left(-e^{i\varphi}\sqrt{r}\right)^l
\left(\sqrt{1-r}\right)^{S-l}
\times{}\\
&{}\times
\left(\sqrt{1-r}\right)^{m-n}
\left(\sqrt{r}\right)^{n-m}
\left(e^{-i\varphi}\right)^{m+n}\times{}\\
&{}\times(-1)^{-m}
\left(a_r^\dagger\right)^{m+n}
\left(a_t^\dagger\right)^{S-m-n}
\ket{0}
\end{aligned}
\\\noalign{\goodbreak}
U\ket{l}_a\ket{S-l}_b={}&
e^{-i\theta \tfrac{S}{2}}
\dfrac{\left(-e^{i\varphi}\sqrt{r}\right)^l\left(\sqrt{1-r}\right)^{S-l}}{\sqrt{l!\,(S-l)!}}
\begin{aligned}[t]
&\sum_{m=0}^l
\sum_{n=0}^{S-l}
\binom{l}{m}
\binom{S-l}{n}
(-1)^{-m}
\times{}\\
&{}\times
\left(e^{-i\varphi}\right)^{m+n}
\left(\sqrt{\tfrac{1-r}{r}}\right)^{m-n}\times{}\\
&{}\times\sqrt{(m+n)!\,(S-m-n)!}\ket{m+n,S-m-n}.
\end{aligned}
\end{align}
Let us substitute $m+n=k$ to change the summation variables. Then,
\begin{align*}
\ket{m+n,S-m-n}={}&\ket{k,S-k}
\end{align*}
and the ranges of $k$ and $m$ are as follows
\begin{gather*}
0\leq m+n=k\leq S\\
0\leq k-m=n\leq S-l\Rightarrow k+l-S\leq m\leq k,
\end{gather*}
\begin{gather*}
\sum_{m=0}^l\sum_{n=0}^{S-l}\Rightarrow \sum_{k=0}^S\kern.5em\sum_{m=\max\{0,k+l-S\}}^{\min\{l,k\}},
\end{gather*}

\begin{align}
U\ket{l}_a\ket{S-l}_b={}&
e^{-i\theta \tfrac{S}{2}}
\dfrac{\left(-e^{i\varphi}\sqrt{r}\right)^l\left(\sqrt{1-r}\right)^{S-l}}{\sqrt{l!\,(S-l)!}}
\begin{aligned}[t]
&\sum_{k=0}^S\kern.5em\sum_{m=\max\{0,k+l-S\}}^{\min\{l,k\}}
\binom{l}{m}
\binom{S-l}{k-m}
(-1)^{-m}
\times{}\\
&{}\times
\left(e^{-i\varphi}\right)^{k}
\left(\sqrt{\tfrac{1-r}{r}}\right)^{2m-k}\sqrt{k!\,(S-k)!}\ket{k,S-k}.
\end{aligned}
\end{align}

The probability amplitude of detecting $k$ and $S-k$ photons behind the beam splitter provided that $l$ and $S-l$ were injected into it is
\begin{align}
\mathcal{A}_S(k,l)={}\bra{k,S-k}U\ket{l,S-l},
\end{align}
thus,
\begin{align}
U\ket{l}_a\ket{S-l}_b={}&
\sum_{k=0}^S \mathcal{A}_S(k,l) \ket{k,S-k},
\end{align}
where
\begin{align}
\mathcal{A}_S(k,l)={}&
e^{-i\theta\tfrac{S}{2}}
\dfrac{\left(-e^{i\varphi}\sqrt{r}\right)^l\left(\sqrt{1-r}\right)^{S-l}}{\sqrt{l!\,(S-l)!}}
\begin{aligned}[t]
&\kern.25em\sum_{m=\max\{0,k+l-S\}}^{\min\{l,k\}}
\binom{l}{m}
\binom{S-l}{k-m}
(-1)^{-m}\left(e^{-i\varphi}\right)^{k}\times{}\\
&{}\times\left(\sqrt{\tfrac{1-r}{r}}\right)^{2m-k}\sqrt{k!\,(S-k)!}
\end{aligned}
\\\noalign{\goodbreak}
{}={}&
e^{-i\theta\tfrac{S}{2}}
\dfrac{\left(-e^{i\varphi}\sqrt{r}\right)^l\left(\sqrt{1-r}\right)^{S-l}}{\sqrt{l!\,(S-l)!}}
\begin{aligned}[t]
&\left(e^{-i\varphi}\right)^{k}
\left(\sqrt{\tfrac{1-r}{r}}\right)^{-k}
\sqrt{k!\,(S-k)!}\times{}\\
&{}\times\sum_{m=\max\{0,k+l-S\}}^{\min\{l,k\}}
\binom{l}{m}
\binom{S-l}{k-m}
(-1)^{-m}
\left(\tfrac{1-r}{r}\right)^{m}.
\end{aligned}
\label{eq:ampl1}
\end{align}
The inner sum over $m$ in Eq.~\ref{eq:ampl1} is a hypergeometric series. In order to simplify it, the identities from Section \ref{Appendix} are used. The four cases below (A-D) correspond to different summation ranges. For simplicity, let us assume that $l\leq S-l$, i.e. $l\leq \tfrac{S}{2}$.
\medskip

\noindent\textbf{Case A:}
$\min\{l,k\}=l$ and $\max\{0,k+l-S\}=0$. This implies $l\leq k\leq S-l$.
\begin{align}
&\kern-4em\sum_{m=0}^{l}
\binom{l}{m}
\binom{S-l}{k-m}
(-1)^{-m}
\left(\tfrac{1-r}{r}\right)^{m}
\\
{}={}&
\sum_{m=0}^{l}
\binom{l}{m}
\binom{S-l}{k-m}
\left(1-\tfrac{1}{r}\right)^{m}
\\
{}={}&
\sum_{m=0}^{l}
\binom{l}{m}
\dfrac{(S-l)!}{(k-m)!\,(S-l-k+m)!}
\dfrac{(S-l-k)!\,k!}{(S-l-k)!\,k!}
\left(1-\tfrac{1}{r}\right)^{m}
\\
{}={}&
\sum_{m=0}^{l}
\binom{l}{m}
\dfrac{(S-l)!}{(S-l-k)!\,k!}
\dfrac{(S-l-k)!}{(S-l-k+m)!}
\dfrac{k!}{(k-m)!}
\left(1-\tfrac{1}{r}\right)^{m}
\\
{}={}&
\sum_{m=0}^{l}
\binom{l}{m}
\binom{S-l}{k}
\dfrac{(-1)^m(-k)_k}{(S-l-k+1)_k}
\left(1-\tfrac{1}{r}\right)^{m}
&\text{cf. (\ref{eq:Pochneg}), (\ref{eq:Pochpos})}
\\\noalign{\goodbreak}
{}={}&
\binom{S-l}{k}
{}_2F_1\left[-l,-k;S-l-k+1;1-\tfrac{1}{r}\right]
&\text{cf. (\ref{eq:2F1fin})}
\\\noalign{\goodbreak}
{}={}&
\binom{S-l}{k}\dfrac{(S-l-k+1+k)_K}{(S-l-k+1)_K}
{}_2F_1\left[-l,-k;-S;\tfrac{1}{r}\right]
&\text{cf. (\ref{eq:DLMF15.8.7})}
\\\noalign{\goodbreak}
{}={}&
\binom{S}{k}
{}_2F_1\left[-l,-k;-S;\tfrac{1}{r}\right].
\end{align}
\medskip

\noindent\textbf{Case B:}
$\min\{l,k\}=k$ and $\max\{0,k+l-S\}=k+l-S$. This implies $S-l\leq k\leq l$, i.e. the empty set.
\medskip

\noindent\textbf{Case C:}
$\min\{l,k\}=k$ and $\max\{0,k+l-S\}=0$. This implies $k\leq l\leq S-l$.
\begin{align}
&\kern-4em\sum_{m=0}^{k}
\binom{l}{m}
\binom{S-l}{k-m}
(-1)^{-m}
\left(\tfrac{1-r}{r}\right)^{m}
\\
{}={}&
\sum_{m=0}^{k}
\binom{l}{m}
\binom{S-l}{k-m}
\left(1-\tfrac{1}{r}\right)^{m}
\\
\noalign{\goodbreak}
{}={}&
\sum_{m=0}^{k}
\dfrac{l!}{m!\,(l-m)!}
\dfrac{(S-l)!}{(k-m)!\,(S-l-k+m)!}
\dfrac{k!\,(S-l-k)!}{k!\,(S-l-k)!}
\left(1-\tfrac{1}{r}\right)^{m}
\\
\noalign{\goodbreak}
{}={}&
\sum_{m=0}^{k}
\dfrac{k!}{m!\,(k-m)!}
\dfrac{(S-l)!}{k!\,(S-l-k)!}
\dfrac{l!}{(l-m)!}
\dfrac{(S-l-k)!}{(S-l-k+m)!}
\left(1-\tfrac{1}{r}\right)^{m}
\\
\noalign{\goodbreak}
{}={}&
\sum_{m=0}^{k}
\binom{k}{m}
\binom{S-l}{k}
\dfrac{(-1)^m(-l)_k}{(S-l-k+1)_k}
\left(1-\tfrac{1}{r}\right)^{m}
&\text{cf. (\ref{eq:Pochneg}), (\ref{eq:Pochpos})}
\\
\noalign{\goodbreak}
{}={}&
\binom{S-l}{k}
{}_2F_1\left[-k,-l;S-l-k+1;1-\tfrac{1}{r}\right]
&\text{cf. (\ref{eq:2F1fin})}
\\
\noalign{\goodbreak}
{}={}&
\binom{S-l}{k}
{}_2F_1\left[-l,-k;S-l-k+1;1-\tfrac{1}{r}\right]
&\text{cf. (\ref{eq:2F1ab})}
\\
\noalign{\goodbreak}
{}={}&
\binom{S-l}{k}\dfrac{(S-l-k+1+k)_K}{(S-l-k+1)_K}
{}_2F_1\left[-l,-k;-S;\tfrac{1}{r}\right]
&\text{cf. (\ref{eq:DLMF15.8.7})}
\\
{}={}&
\binom{S}{k}
{}_2F_1\left[-l,-k;-S;\tfrac{1}{r}\right]
\end{align}

\goodbreak

\noindent\textbf{Case D:}
$\min\{l,k\}=l$ and $\max\{0,k+l-S\}=k+l-S$. This implies $l\leq S-l\leq k$. To compute the sum, the following substitution is used: $m=l-m'$.
\begin{align}
&\sum_{m=k+l-S}^{l}
\binom{l}{m}
\binom{S-l}{k-m}
\left(1-\tfrac{1}{r}\right)^{m}
=
\sum_{m'=0}^{S-k}
\binom{l}{l-m'}
\binom{S-l}{k+m'-l}
\left(1-\tfrac{1}{r}\right)^{l-m'}
\\\noalign{\goodbreak}
{}={}&
\left(1-\tfrac{1}{r}\right)^{l}
\begin{aligned}[t]
&\sum_{m'=0}^{S-k}
\binom{S-k}{m'}
\dfrac{m'!\,(S-k-m')!}{(S-k)!}
\dfrac{l!}{(l-m')!\,m'!}\times{}\\
&{}\times\dfrac{(S-l)!}{(S-k-m')!\,(k+m'-l)!}
(-1)^{-m'}\left(\tfrac{1}{r}-1\right)^{-m'}
\end{aligned}
\\\noalign{\goodbreak}
{}={}&
\left(1-\tfrac{1}{r}\right)^{l}
\sum_{m'=0}^{S-k}
\binom{S-k}{m'}
(-1)^{m'}
\dfrac{l!}{(l-m')!}
\dfrac{(S-l)!}{(S-k)!\,(k+m'-l)!}
\left(\tfrac{1}{r}-1\right)^{-m'}
\\\noalign{\goodbreak}
{}={}&
\left(1-\tfrac{1}{r}\right)^{l}
\begin{aligned}[t]
&\sum_{m'=0}^{S-k}
\binom{S-k}{m'}
(-1)^{m'}(-1)^{m'}(-l)_{m'}\times{}\\
&{}\times\dfrac{(S-l)!}{(S-k)!\,(k-l+m')!}
\dfrac{(k-l)!}{(k-l)!}
\left(\tfrac{r}{1-r}\right)^{m'}
\end{aligned}
\\\noalign{\goodbreak}
{}={}&
\left(1-\tfrac{1}{r}\right)^{l}
\begin{aligned}[t]
&\sum_{m'=0}^{S-k}
\binom{S-k}{m'}
(-1)^{m'}(-l)_{m'}
\dfrac{(S-l)!}{(S-k)!\,(S-l-S+k)!}\times{}\\
&{}\times\dfrac{(k-l)!}{(k-l+m')!}
\left(\tfrac{r}{r-1}\right)^{m'}
\end{aligned}
&\text{cf. (\ref{eq:Pochneg}), (\ref{eq:Pochpos})}
\\
\noalign{\goodbreak}
{}={}&
\left(1-\tfrac{1}{r}\right)^{l}
\binom{S-l}{S-k}
\sum_{m'=0}^{S-k}
\binom{S-k}{m'}
(-1)^{m'}\dfrac{(-l)_{m'}}{(k-l+1)_{m'}}
\left(\tfrac{r}{r-1}\right)^{m'}
&\text{cf. (\ref{eq:2F1fin})}
\\\noalign{\goodbreak}
{}={}&
\left(1-\tfrac{1}{r}\right)^{l}
\binom{S-l}{S-k}
{}_2F_1\left[-(S-k),-l;k-l+1;\tfrac{r}{r-1}\right]
&\text{cf. (\ref{eq:DLMF15.8.7})}
\\\noalign{\goodbreak}
{}={}&
\left(1-\tfrac{1}{r}\right)^{l}
\begin{aligned}[t]
&\binom{S-l}{S-k}
\dfrac{(k-l+1+l)_{S-k}}{(k-l+1)_{S-k}}\times{}\\
&{}\times{}_2F_1\left[-(S-k),-l;1-S+k-l-(k-l+1);\tfrac{-1}{r-1}\right]
\end{aligned}
\\\noalign{\goodbreak}
{}={}&
\left(1-\tfrac{1}{r}\right)^{l}
\binom{S}{S-k}
{}_2F_1\left[-(S-k),-l;-S;\tfrac{-1}{r-1}\right]
&\text{cf. (\ref{eq:2F1Pfaff})}
\\\noalign{\goodbreak}
{}={}&
\underbrace{\left(1-\tfrac{1}{r}\right)^{l}
	\left(1-\tfrac{-1}{r-1}\right)^{l}}_{=1}
\binom{S}{k}
{}_2F_1\left[-S+(S-k),-l;-S;\dfrac{\tfrac{-1}{r-1}}{\tfrac{-1}{r-1}-1}\right]
\\\noalign{\goodbreak}
{}={}&
\binom{S}{k}
{}_2F_1\left[-k,-l;-S;\tfrac{1}{r}\right]
&\text{cf. (\ref{eq:2F1ab})}
\\\noalign{\goodbreak}
{}={}&
\binom{S}{k}
{}_2F_1\left[-l,-k;-S;\tfrac{1}{r}\right].
\end{align}

\noindent
Summarizing, the inner sum in Eq.~\ref{eq:ampl1} equals $\binom{S}{k}{}_2F_1\left[-l,-k;-S;\tfrac{1}{r}\right]$ under the assumption that $l\leq \tfrac{S}{2}$. The probability amplitude can be rewritten into the following form
\begin{align}
\mathcal{A}_S(k,l)={}&
\dfrac{\left(-e^{i\varphi}\sqrt{r}\right)^l\left(\sqrt{1-r}\right)^{S-l}}{\sqrt{l!\,(S-l)!}}
e^{-i\theta \tfrac{S}{2}}
\left(e^{-i\varphi}\right)^{k}
\left(\sqrt{\tfrac{1-r}{r}}\right)^{-k}
\sqrt{k!\,(S-k)!}\,
\binom{S}{k}{}_2F_1\left[-l,-k;-S;\tfrac{1}{r}\right],
\end{align}

\begin{align*}
\sqrt{\dfrac{k!\,(S-k)!}{l!\,(S-l)!}}\binom{S}{k}={}&
\sqrt{\binom{S}{k}\binom{S}{l}},
\end{align*}

\begin{align}
\mathcal{A}_S(k,l)={}&
\sqrt{\binom{S}{k}\binom{S}{l}}
(-1)^{l}
\left(e^{i\varphi}\right)^{l-k}
e^{-i\theta \tfrac{S}{2}}
\left(\sqrt{1-r}\right)^S
\left(\sqrt{\tfrac{r}{1-r}}\right)^{l+k}
{}_2F_1\left[-l,-k;-S;\tfrac{1}{r}\right]
\\
{}={}&
\sqrt{\binom{S}{k}\binom{S}{l}}
(-1)^{l}
\left(e^{i\varphi}\right)^{l-k}
e^{-i\theta \tfrac{S}{2}}
\left(\cos\tfrac{\theta}{2}\right)^S
\left(\tan\tfrac{\theta}{2}\right)^{l+k}
{}_2F_1\left[-l,-k;-S;\left(\sin\tfrac{\theta}{2}\right)^{-2}\right]
\\
{}={}&
\sqrt{\binom{S}{k}\binom{S}{l}}
(-1)^{l}
\left(e^{i\varphi}\right)^{l-k}
e^{-i\theta \tfrac{S}{2}}
\left(\cos\tfrac{\theta}{2}\right)^S
\left(\tan\tfrac{\theta}{2}\right)^{l+k}
{}_2F_1\left[-k,-l;-S;\left(\sin\tfrac{\theta}{2}\right)^{-2}\right].
\label{eq:finalA}
\end{align}
The photon number statistics behind the beam splitter is given by the probability $p_{S}(k,l)= \lvert \mathcal{A}_S(k,l) \rvert^2$
\begin{align}
p_{S}(k,l)={}&
\binom{S}{k}\binom{S}{l}
\left(\cos\tfrac{\theta}{2}\right)^{2S}
\left(\tan\tfrac{\theta}{2}\right)^{2(l+k)}
\left\lvert{}_2F_1\left[-l,-k;-S;\left(\sin\tfrac{\theta}{2}\right)^{-2}\right]\right\rvert^2
= p_{S}(l,k).
\label{eq:p_Skl}
\end{align}

\subsection{Kravchuk transform}

The $\alpha$-fractional Kravchuk transform of an input sequence $x_n=f(\xi_{n})$, where $n=0,1,\ldots,N$ and $\xi_n=(n-N/2)$, is defined as follows\cite{Atakishiyev1997} (cf.\ Eq.~5.2)
\begin{align}
\mathbf{X}_n={}\sum_{n'=0}^N F_{n,n'}^{\alpha}\,x_{n'},
\end{align}
\begin{align}
F_{n,n'}^{\alpha}={}& e^{i\tfrac{\pi}{2}(n+n'-N\alpha/2)} \sqrt{\binom{N}{n}\binom{N}{n'}} \cos^N\left(\tfrac{\pi\alpha}{4}\right)\tan^{n+n'}\left(\tfrac{\pi\alpha}{4}\right)\,{}_2F_1\left[-n,-n';-N;\sin^{-2}\left(\tfrac{\pi\alpha}{4}\right)\right]
=
F_{n',n}^{\alpha}
\\
={}& 
e^{i\tfrac{\pi}{2}(n'-n-N\alpha/2)} \sqrt{\dfrac{n!\,(N-n)!}{n'!\,(N-n')!}} \sin^{n'-n}\left(\tfrac{\pi\alpha}{4}\right) \cos^{N-n'-n}\left(\tfrac{\pi\alpha}{4}\right)\,k_n^{[\sin^{2}(\pi\alpha/4)]} (n',N)
\\
={}& 
e^{i\tfrac{\pi}{2}(n'-n-N\alpha/2)}
\phi_n^{(p)}(n'-Np,N),
\label{eq:Kravchuk_short}
\end{align}
where $k_n^{(p)}(n',N)$ is a Kravchuk polynomial and $\phi_n^{(p)}(n'-Np,N)$ is a Kravchuk function.

We used the following relations\cite{Atakishiev1994}
\begin{align}
k_n^{(p)}(n', N)={}& (-1)^n \, \binom{N}{n}\, p^n\;{}_2F_1\left[-n, -n'; -N; \tfrac{1}{p}\right],\\
\phi_n^{(p)}(n'-Np,N)={}&
\sqrt{\dfrac{n!\,(N-n)!}{n'!\,(N-n')!}} \,  \sqrt{p^{n'-n}(1-p)^{N-n-n'}}\;  k_n^{(p)}(n',N),\\
\phi_n^{(p)}(n'-Np,N)={}&
(-1)^{n+n'}\,\phi_{n'}^{(p)}(n-Np,N),
\end{align}
as well as the fact that the Kravchuk functions are orthonormal
\begin{align}
\sum_{n'=0}^N \phi_n^{(p)}(n'-Np,N)\, {}& \phi_m^{(p)}(n'-Np,N) = \delta_{n,m}.
\end{align}
\medskip

\noindent
Now we turn $\mathcal{A}_S(k,l)$ shown in Eq.~\ref{eq:finalA}
to the form of Eq.~\ref{eq:Kravchuk_short}
\begin{equation}
\boxed{%
	\begin{aligned}
	\mathcal{A}_S(k,l)
	={}& e^{-i \theta \tfrac{S}{2}}\, e^{i\varphi (l-k)}\, (-1)^{k+l}\;
	\phi_k^{\left(r\right)}(l-S r, S) \\
	={}& e^{i\tfrac{\pi}{2}\left(\tfrac{2(\pi + \varphi)}{\pi}(l-k) - S \tfrac{\theta}{\pi}\right)} \;
	\phi_k^{\left(r\right)}(l-S r, S)\\
	={}& e^{-i \theta \tfrac{S}{2}}\, e^{i\varphi (l-k)}\;
	\phi_l^{\left(r\right)}(k-S r, S) \\
	={}& e^{i\tfrac{\pi}{2}\left(\tfrac{2\varphi}{\pi}(l-k) - S \tfrac{\theta}{\pi}\right)} \;
	\phi_l^{\left(r\right)}(k-S r, S),
	\end{aligned}
}
\end{equation}
where $r=\sin^{2}\tfrac{\theta}{2}$.

In specific, if we take $\varphi=-\tfrac{\pi}{2}$ and rearrange terms
\begin{align}
\mathcal{A}_S(k,l)
{}={}&
e^{i \tfrac{\pi}{2} (k+l - S \tfrac{ \theta}{\pi})}
\sqrt{\binom{S}{k}\binom{S}{l}}
\left(\cos\tfrac{\theta}{2}\right)^S
\left(\tan\tfrac{\theta}{2}\right)^{l+k}
{}_2F_1\left[-k,-l;-S;\left(\sin\tfrac{\theta}{2}\right)^{-2}\right]\\
{}={}& F_{k,l}^{\tfrac{2\theta}{\pi}}\\
{}={}&
e^{i\tfrac{\pi}{2}(l-k-S \tfrac{\theta}{\pi})}
\phi_k^{\left(\sin^{2}\tfrac{\theta}{2}\right)}(l-S \sin^{2}\tfrac{\theta}{2}, S),
\end{align}
\begin{equation}
\boxed{%
	\mathcal{A}_S(k,l)
	=
	e^{i\tfrac{\pi}{2}(l-k-S \tfrac{\theta}{\pi})}
	\phi_k^{\left(r\right)}(l-S r, S).
}
\end{equation}

\subsection{Quantum Kravchuk transform on a beam splitter}

Let us send a superposition $\sum_{l=0}^S x_l \ket{l,S-l}$ to a BS. The superposition amplitudes encode the sequence $(x_1, \dots, x_S)$ to be transformed. We will compute the probabilities of detecting $\ket{k}$ and $\ket{S-k}$ photons behind the BS
\begin{align}
\left\lvert\langle k,S-k\rvert U_0\,U_{BS}\left(\sum_{l=0}^S x_l\cdot \lvert l,S-l\rangle\right) \right\rvert^2
{}={}&
\left\lvert\sum_{l=0}^S x_l\cdot
\langle k,S-k\rvert U_0\,U_{BS}\lvert l,S-l\rangle \right\rvert^2
\\\noalign{\goodbreak}
{}={}&
\left\lvert\sum_{l=0}^S x_l\cdot e^{-i\theta\tfrac{S}{2}}
\langle k,S-k\rvert U_{BS}\lvert l,S-l\rangle \right\rvert^2
\\\noalign{\goodbreak}{}={}&
\left\lvert\sum_{l=0}^S x_l\cdot
\mathcal{A}_S^{(r)}(k,l) \right\rvert^2
\\\noalign{\goodbreak}{}={}&
\left\lvert\sum_{l=0}^S x_l\cdot
e^{-i\theta\tfrac{S}{2}}
e^{i\tfrac{\pi}{2}(l-k)}
\phi_k^{\left(r\right)}(l-S r, S)\right\rvert^2
\\
{}={}&
\left\lvert \mathrm{X}_k \right\rvert^2.
\end{align}

It is clear now that multi-photon interference on a  beam splitter followed by photon-counting detection implements $\alpha = \tfrac{2 \theta}{\pi}$-fractional QKT of the input probability amplitudes
\begin{align}
(x_0, x_1, \ldots, x_{S}) \to (\lvert \mathrm{X}_0 \rvert^2, \lvert \mathrm{X}_1 \rvert^2, \dots, \lvert \mathrm{X}_S \rvert^2),
\end{align}
where $\lvert \mathrm{X}_k \rvert^2$ are experimentally determined photon number statistics for $k=0, \ldots, S$.

\section{Gauss hypergeometric function}
\label{Appendix}

{\bf Definition.} The Gauss hypergeometric function is a special function defined with the following hypergeometric series
\begin{align}
{}_2F_1(a,b;c;z)={}&
\sum_{k=0}^{\infty} \dfrac{(a)_k\,(b)_k}{(c)_k}\,\dfrac{z^k}{k!},
\label{eq:2F1def}
\end{align}
where $a$, $b$ and $c$ are parameters, $z$ is an argument and $(x)_k$ is the Pochhammer symbol
\begin{align}
(x)_k={}x(x+1)(x+2)\cdots(x+k-1).
\end{align}
In general, all ${}_2F_1$ arguments and the parameter may be complex, $a, b, c, z\in \mathbb{C}$ however, within this note the arguments are always integer, $a, b, c\in\mathbb{Z}$ and the parameter is real, $z\in\mathbb{R}$.

\noindent
{\bf Properties.} 
The Pochhammer symbol can be expressed as a division of factorials
\begin{align}
\dfrac{a!}{(a-k)!}=&(-1)^k(-a)_k,\label{eq:Pochneg}\\
\dfrac{a!}{(a+k)!}=&\dfrac{1}{(a+1)_k}.\label{eq:Pochpos}
\end{align}
The form of Eq.~\ref{eq:2F1def} implies that the arguments $a$ and $b$ can be swapped
\begin{align}
{}_2F_1(a,b;c;z)={}&
\sum_{k=0}^{\infty} \dfrac{(a)_k\,(b)_k}{(c)_k}\,\dfrac{z^k}{k!}
=
\sum_{k=0}^{\infty} \dfrac{(b)_k\,(a)_k}{(c)_k}\,\dfrac{z^k}{k!}
=
{}_2F_1(b,a;c;z).
\label{eq:2F1ab}
\end{align}
In case of a negative $a$ or $b$, the infinite sum in Eq.~\ref{eq:2F1def} is truncated because $(x)_k=0$ if $x$ is a negative integer and $k>-x$. Let us assume that $a<0$ and $b\geq 0\vee b<a$. Then, let $m=-a$
\begin{align}
{}_2F_1(-m,b;c;z)={}&
\sum_{k=0}^{\infty} \dfrac{(-m)_k\,(b)_k}{(c)_k}\,\dfrac{z^k}{k!}
&\text{cf. (\ref{eq:Pochneg})}
\nonumber\\
{}={}&
\sum_{k=0}^{m} (-1)^k\dfrac{m!}{k!\,(m-k)!}\,\dfrac{(b)_k}{(c)_k}\,z^k
\nonumber\\
{}={}&
\sum_{k=0}^m \binom{m}{k}(-1)^k\dfrac{(b)_k}{(c)_k}z^k.\label{eq:2F1fin}
\end{align}
Moreover, for the same assumptions as in case of Eq.~\ref{eq:2F1fin}, the following transformation can be used to change $z$ to $1-z$ [NIST Digital Library of Mathematical Functions, 15.8.7]
\begin{align}
{}_2F_1(-m,b;c;z)={}&\dfrac{(c-b)_m}{(c)_m}{}_2F_1(-m,b;b-c-m+1;1-z).
\label{eq:DLMF15.8.7}
\end{align}
Identities analogous to Eqs.~\ref{eq:2F1fin} and \ref{eq:DLMF15.8.7} are also valid for negative $b$ and $a\geq 0\vee a<b$, due to Eq.~\ref{eq:2F1ab}.
\medskip

\noindent
Finally, the following Pfaff's hypergeometric transformation is valid for any $a, b, c$ and $z$
\begin{align}
{}_2F_1(a,b;c;z)={}&(1-z)^{-b}{}_2F_1(c-a,b;c;z/(z-1)).\label{eq:2F1Pfaff}
\end{align}

\section{Characterization of the setup}
\label{characterization_setup}

In order to estimate transmission losses, we performed Klyshko efficiency measurements on the setup. In a Klyshko measurement with one SPDC source and binary detectors, one counts single events $C_A$, $C_B$ from either output channel and coincidence clicks $C_{AB}$ between both channels and defines the Klyshko efficiencies $\eta_A$ and $\eta_B$
\begin{equation}
\eta_B = \frac{C_{AB}}{C_A}
\end{equation}
and vice versa. For low pump powers, these Klyshko efficiencies show a linear pump power dependency, and their intercept is a measure at zero pump power of total transmission efficiency (including both propagation and detection losses) of the associated spatial mode\cite{Klyshko1980}.

We pumped each of our SPDC sources, one at a time, with the variable beam-splitter in position $50:50$, at successively lower power values. The resulting four-mode correlated photon statistics  were then transformed into binary ``photon(s)/no-photon'' datasets to emulate standard binary detectors such as avalanche photo-diodes, and we determined the total efficiencies of  the heralding modes to be $\eta_1=50.3\%$  and $\eta_4=48.5\%$. The beam-splitter modes, carrying each a $3\kern.25em\mathrm{dB}$ loss from the splitter itself and an additional $1\kern.25em\mathrm{dB}$ due to splitter insertion loss and fiber-to-fiber coupling loss, exhibit a total efficiency of $\eta_2=21.6\%$  and $\eta_3=20.6\%$. Taking into account the additional optical elements in the splitter modes, the efficiencies are consistent. We account for the transmission losses of approximately $50\% \approx 3\kern.25em\mathrm{dB}$ with $1\kern.25em\mathrm{dB}$ initial fiber in-coupling loss due to spatial mode mismatch, $0.25\kern.25em\mathrm{dB}$ from imperfect detectors, and the rest from three FC/PC fiber-to-fiber couplers per mode as well as bending losses in the transmission fibers between the experimental setup and the detectors.

\begin{figure}\centering
	\includegraphics[height=6cm]{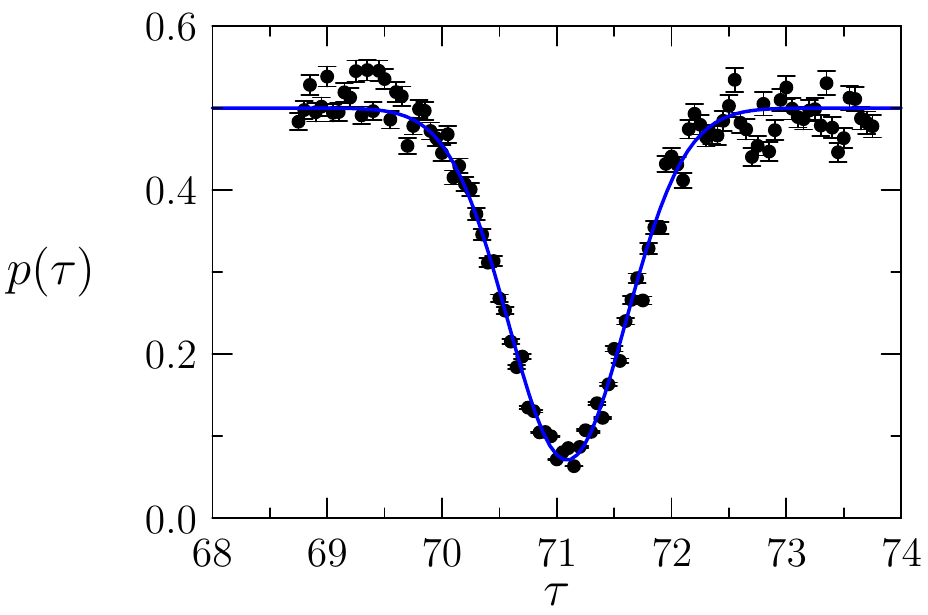}
	\caption{\textbf{HOM dip.} Black dots with error bars represent the experimental results whereas the blue line is a fitted curve. The maximal visibility amounts to $85.9\%$ which proves the quantum nature of impinging multiphoton states.}
	\label{fig:hom}
\end{figure}

Fig.~\ref{fig:hom} shows the standard HOM interference dip between both sources measured with binary detectors (InGaAs APDs) for a small mean photon number of the order of $10^{-4}$ in order to test the setup. The maximal visibility achieved is $V_\textrm{HOM}=85.9\%$. An independent measurement of the second order correlation function for each SPDC source $g^{(2)}=\frac{\braket{n^2}-\braket{n}}{\braket{n}^2}\geq1.86\approx1+V_\textrm{HOM}$ is consistent with this result. From this, we can  infer an effective Schmidt mode number of $K=\frac{1}{g^{(2)}-1}=1.16$, \cite{Christ2011} i.e.\ both of our SPDC sources are close to being single-mode.

The TES detectors used in the experiment were thoroughly characterized with quantum tomography methods\cite{Humphreys2015}. Their quantum efficiency is above $90\%$.

\section{Analysis of the experimental data}

\subsection{HOM visibilities}

The second-order visibility exceeding the classical value of $50\%$ certifies quantum nature of the HOM interference and thus, the fractional QKT. The visibility is computed with the following formula\cite{Filip2002}
\begin{equation}
v^{(2)} = \dfrac{n_{\text{max}} - n_{\text{min}}}{n_{\text{max}} + n_{\text{min}}},
\end{equation}
where $n_{\text{max}}$ and $n_{\text{min}}$ are the maximal and minimal number of events registered by the TES detectors for the given photon number $S$.  

The obtained values are gathered in Tab.~\ref{tab:visibilities}.  For $S=5$ it was always greater than $50\%$. The visibility of interference of $\lvert 1,1\rangle$ given in this Table is much lower than the one reported in Fig.~\ref{fig:hom}. This is because in order to perform quantum simulations with $S>2$ we increased the power of a laser pumping our source. However, this power increase is too small to affect the analysis from Section~\ref{characterization_setup}, i.e.\ the source operates in the parametric regime and our photon-number states are near single-mode.

\begin{table*}\centering
	\begin{minipage}[t]{12cm}\centering
		\begin{tabular}{lcccc}\hline
			$\lvert\psi\rangle$& $r=0.05$& $r=0.3$& $r=0.5$& $r=0.95$\\\hline\hline
			$\lvert 0,1\rangle$& $87.2\%\pm0.1\%$& $35.1\%\pm0.1\%$& $0.8\%\pm0.0\%$& $87.8\%\pm0.1\%$\\
			& $(\bar{n}=0.2134)$& $(\bar{n}=0.2082)$& $(\bar{n}=0.2097)$& $(\bar{n}=0.2082)$\\\hline
			$\lvert 0,2\rangle$& $98.2\%\pm0.3\%$& $59.9\%\pm0.2\%$& $26.5\%\pm0.1\%$& $99.0\%\pm0.3\%$\\
			& $(\bar{n}=0.2134)$& $(\bar{n}=0.2082)$& $(\bar{n}=0.2097)$& $(\bar{n}=0.2082)$\\\hline
			$\lvert 0,3\rangle$& $99.7\%\pm0.8\%$& $78.6\%\pm0.7\%$& $52.4\%\pm0.4\%$& $99.9\%\pm0.8\%$\\
			& $(\bar{n}=0.2134)$& $(\bar{n}=0.2082)$& $(\bar{n}=0.1996)$& $(\bar{n}=0.2082)$\\\hline
			$\lvert 0,4\rangle$& $99.1\%\pm2.5\%$& $87.6\%\pm2.2\%$& $65.7\%\pm1.7\%$& $99.9\%\pm2.5\%$\\
			& $(\bar{n}=0.2134)$& $(\bar{n}=0.2082)$& $(\bar{n}=0.2097)$& $(\bar{n}=0.2082)$\\\hline
			$\lvert 0,5\rangle$& $97.8\%\pm6.2\%$& $96.7\%\pm7.2\%$& $71.4\%\pm4.6\%$& $98.6\%\pm7.2\%$\\
			& $(\bar{n}=0.2076)$& $(\bar{n}=0.2082)$& $(\bar{n}=0.2097)$& $(\bar{n}=0.1983)$\\\hline
			$\lvert 1,2\rangle$& $74.8\%\pm0.8\%$& $18.9\%\pm0.3\%$& $50.3\%\pm0.2\%$& $79.3\pm0.8\%\%$\\
			& $(\bar{n}=0.2043)$& $(\bar{n}=0.2082)$& $(\bar{n}=0.1997)$& $(\bar{n}=0.2051)$\\\hline
			$\lvert 2,2\rangle$& $94.5\%\pm2.2\%$& $42.5\%\pm1.0\%$& $50.6\%\pm1.2\%$& $93.8\%\pm2.3\%$\\
			& $(\bar{n}=0.2088)$& $(\bar{n}=0.2141)$& $(\bar{n}=0.2097)$& $(\bar{n}=0.2051)$\\\hline
			$\lvert 2,3\rangle$& $97.7\%\pm7.0\%$& $76.6\%\pm4.7\%$& $54.8\%\pm3.7\%$& $99.5\%\pm7.5\%$\\
			& $(\bar{n}=0.2043)$& $(\bar{n}=0.2150)$& $(\bar{n}=0.2097)$& $(\bar{n}=0.1969)$\\\hline\hline
		\end{tabular}
	\end{minipage}
	\caption{\textbf{Second-order interferometric visibilities in HOM interference.} The visibility above $50\%$ proves quantum character of the interference. Two-mode Fock states $\lvert\psi\rangle$ impinging on a beam splitter of a variable reflectivity $r$ implement the fractional QKTs. $\bar{n}$ denotes the mean number of interfering photons reached in the experiment.}
	\label{tab:visibilities}
\end{table*}

\subsection{Computation of probability distributions and estimation of errors}

Experimental demonstration of two-mode multi-photon HOM interference requires collecting photon-number statistics, which are then compared with theoretical probability distributions. The statistics result from multiple measurements performed with the setup depicted in Fig.~1\textbf{B} in the main text. The heralding modes (A \& D) inform about the input state fed into the variable BS and together with the output modes are measured by highly efficient photon counting TES detectors. Thus, each measurement results in a 4-tuple consisting of the number of photons registered by TES${}_{1-4}$, denoted as $(n_1, n_2, n_3, n_4)$ and corresponding to photon-number states in modes A--D\cite{Gerrits2011}. In a single run, the SPDC source produces input Fock states consisting of up to approximately 10 photons with probability governed by the pump power (see the \textit{Materials and Methods} section in the main text). The detectors register all possible values of $n_i\in[0,10]$, $i=1,\dots,4$. The automation software stores this data in a database and assigns the number of events to each possible tuple. During a single 400-second run, approx.~$10^9$ data points are collected.

In order to obtain a photon-number statistics for a given $r=\sin^2\tfrac{\theta}{2}$ and input Fock state a post-processing is required. The database is searched for a given pair $(n_1, n_4)$ which determines the two-mode Fock state at the BS input. Then, only records fulfilling the condition $n_1+n_4=n_2+n_3$ are selected as they may correspond to the case of no losses in all paths. For the given $(n_1, n_4)$ the individual probabilities are computed as
\begin{displaymath}
p_S(k,n_1+n_4-k) = \dfrac{N\left(n_1,k,n_1+n_4-k,n_4\right)}{S(n_1,n_4)},
\end{displaymath}
where $N(n_1,n_2,n_3,n_4)$ denotes the number of events of registering the given 4-tuple, $S(n_1,n_4) = \sum_{m=0}^{n_1+n_4} N(n_1,m,n_1+n_4-m,n_4)$ is the total number of contributing data points and $k$ as well as $n_1+n_4-k$ are the photon numbers registered at the BS outputs. The full probability distribution consists of $n_1+n_4+1$ values for $k$ ranging from 0 to $n_1+n_4$.

For the TES detectors, due to the overlap between the outcomes associated with neighboring photon numbers, an $\lvert n\rangle$ state results in a value of $n\pm 1$, where $n$ is registered with probability over $0.9$ and the probabilities of $n-1$ and $n+1$ are below $0.1$ with $p(n-1)\gg p(n+1)$. Therefore, the absolute error of a single measurement $\Delta n=\pm 1$. As the computation of probability is based on $S(n_1,n_4)$ data points, the measurement uncertainty equals
\begin{displaymath}
\Delta p = \dfrac{\lvert\Delta n\rvert}{\sqrt{S(n_1,n_4)}}\approx \dfrac{1}{\sqrt{S(n_1,n_4)}}.
\end{displaymath}
The data post-processing and error estimation was done with a Python script, which prepared input files for the Asymptote plotting software.  The probability distributions for an ideal system were computed with Eq.~\ref{eq:p_Skl}.  Factorials and binomial coefficients were approximated with the standard \verb|lgamma(n)| function.

\subsection{Realistic theoretical model}

Actual experimental results (Fig.~3 in the main text) were compared with an enhanced realistic theoretical model which allowed to assess the imperfections of the system.  The model includes the following parameters: average photon numbers at the outputs of both SPDCs, strength of the fiber coupling, losses in heralded and interfering modes as well as efficiencies of individual TES detectors.

The computations are done with $6\times 6$ complex matrices, where the indexes 1-2 correspond to heralded modes and 3-4 to the outputs of the variable beam splitter. The indexes 5-6 are responsible for the losses in modes entering the beam splitter, which are modeled by two additional beam splitters which bring the SPDC outputs B and C to interference with the vacuum state. The TES detectors are described by the probability of detecting $n_d$ photons in a Fock state $\lvert n_{\text{in}}\rangle$, given by the following formula
\begin{equation}
p_{\text{TES}}(n_{\text{in}}, n_d, \eta) = \begin{cases}
\dbinom{n_{\text{in}}}{n_d} (1-\eta)^{n_{\text{in}}-n_d}\,\eta^{n_d}& 
\text{if $n_d\leq n_{\text{in}}$,}\\
0& \text{otherwise,}
\end{cases}
\label{TES_distr}
\end{equation}
where $\eta$ is the efficiency of the detector, additionally decreased to model imperfections in optical signal transfer (e.g.\ fiber coupling). The distribution in Eq.~\ref{TES_distr} well models detectors used in the experiment\cite{Humphreys2015}.

\begin{figure}\centering
	\includegraphics[width=15cm]{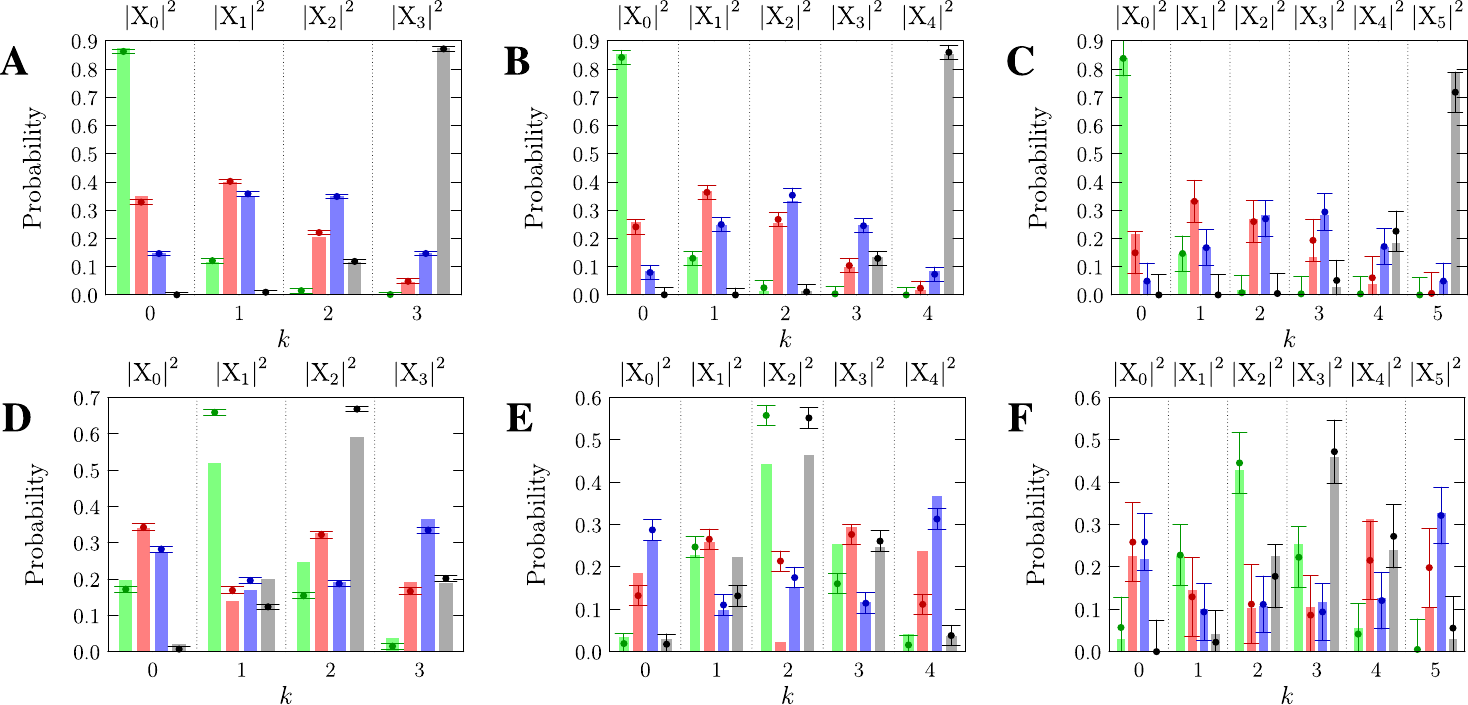}
	\caption{\textbf{Photon number statistics resulting from  Fock state $\ket{l,S-l}$ interference.} (\textbf{A}) $\lvert 0, 3\rangle$, (\textbf{B}) $\lvert 0, 4\rangle$, (\textbf{C}) $\lvert 0,5\rangle$, (\textbf{D}) $\lvert 1, 2\rangle$, (\textbf{E}) $\lvert 2, 2 \rangle$, (\textbf{F}) $\lvert 2,3\rangle$. The BS reflectivities are $r=0.05$ (green), $0.2$ (red), $0.5$ (blue) and $0.95$ (gray). Vertical bars represent theoretical values computed for a realistic system, while dots are values determined in experiment -- the probabilities of detecting $\ket{k}$ and $\ket{S-k}$ photons behind the BS. The parameters of computation: mean number of photons generated by SPDC equal to $0.2$, TES detection efficiency -- $0.9$, fiber coupling -- $0.7$ and overall losses in the system -- $50\%$. The states (\textbf{A})-(\textbf{C}) encode sequences $(x_0=1, x_1=0, \dots, x_S=0)$, while in (\textbf{D}) -- $(0, 1, 0, 0)$, (\textbf{E}) -- $(0, 0, 1, 0, 0)$, (\textbf{F}) -- $(0, 0, 1, 0, 0, 0)$, respectively. The measured probabilities set their QKTs $(\lvert \mathrm{X}_0 \rvert^2, \lvert \mathrm{X}_1 \rvert^2, \dots, \lvert \mathrm{X}_S \rvert^2)$, $\lvert X_k\rvert^2=\lvert \sum_{l=0}^S \mathcal{A}^{(r)}_S(k,l) \cdot x_l \rvert^2$ of fractionality $\alpha = 0.28$ (green), $0.60$ (red), $1.00$ (blue) and $1.72$ (gray). }
	\label{fig:realistic}
\end{figure}

The numerical program was written in the Java programming language and run on a standard PC. It allows to compute output probability distributions $p_S(k,l)$ for given set of model parameters and given readouts at heralded modes $(n_1,n_4)$. The computation results were passed to Python scripts which prepared Asymptote data files to be merged with experimental plots. The computations were performed for the same input Fock states as in Fig.~3 in the main text and mean number of photons equal to $0.2$. Then, the program was run for various parameters in order to fit the theoretical distributions to the actual experimental data. The results are presented in Fig.~\ref{fig:realistic}.

\section{Mapping between qudit and interacting spin-$\tfrac{1}{2}$-chain quantum computer architectures}

Any state of a $d$-level qudit can be encoded in a chain of $d$ qubits where only one qubit is excited at a time, i.e. using the single excitation basis $\ket{1,0,\dots, 0}$, $\ket{0,1,\dots, 0}$, etc. The XY Heisenberg model maps the next-neighbor interaction in the chain to the qudit rotation discussed in the main text.

\subsection{XY model}

Let us consider an interacting chain of $N$ qubits governed by the following Hamiltonian
\begin{equation}
H_{XY}=\sum_{n=1}^N \dfrac{J_n}{2} \left[\sigma_n^x\,\sigma_{n+1}^x+\sigma_n^y\,\sigma_{n+1}^y\right],
\end{equation}
where $\sigma_n^x$, $\sigma_n^y$, $\sigma_n^z$ are the Pauli operators acting on the $n$th qubit and $J_n$ denote couplings between neighboring qubits in the chain. 

We first note that a spin-$\tfrac{N-1}{2}$ particle corresponds to an $N$-qubit chain with relabeled basis vectors as $\ket{m}$, where $m=-\tfrac{N-1}{2} +n-1$\cite{Christandl2004}. The $N$-qubit Hilbert space is of dimension $2^N$. Let us restrict $H_{XY}$ to the N-dimensional single-excitation subspace of this system. This subspace is spanned by the basis vectors $\ket{n}$, $n=1,\dots, N$, corresponding to spin configurations in which all spins are ``down'' apart from just one spin at the vertex $n$ which is ``up'', i.e. by the eigenstates of the $\sigma^z_{tot} = \sum_i \sigma_i^z$ operator. Then $H_{XY}$ is identical to the Hamiltonian of a spin-$\tfrac{N-1}{2}$ particle $H=\lambda S_x$, where $\lambda$ is a constant. Here  $J_n=\tfrac{\lambda}{2}\sqrt{n(N-n)}$. 
This particular form of $J_n$ allows us to link the XY with the BS interaction. The BS infinitesimal evolution turn  the input state $\ket{l, S-l}$ into the superposition
\begin{equation}
H_{BS}\!\ket{l, S-l} = q_{l,l-1} \ket{l-1, S-l+1} + q_{l,l+1}\ket{l+1, S-l-1},
\end{equation}
with the amplitudes 
\begin{equation}
q_{l, l+1} = \tfrac{\sqrt{(l+1)(S-l)}}{2}.
\label{amplitudes}
\end{equation}
The amplitudes reproduce $J_n$ for $N=S+1$, $n=l+1$ and $\lambda=1$.

\subsection{Example: quantum annealing processor}

A Hamiltonian describing quantum annealing processor based on $N$ interacting  qubits reads
\begin{equation}
H_S(s) = \mathcal{E}(s)H_P-\dfrac{1}{2}\sum_{i}\Delta(s)\sigma_i^x, \quad i=1,\dots,N,
\end{equation}
where $s$ denotes time ($s=t/t_f$, $t\in[0,t_f]$), $\mathcal{E}(s)$ and $\Delta(s)$ are the transverse and longitudinal energies, respectively. $H_P$ is a dimensionless Hamiltonian
\begin{equation}
H_P=-\sum_{i} h_i\,\sigma_i^z + \sum_{i<j} J_{ij}\,\sigma_i^z\,\sigma_j^z,
\end{equation}
where biases $h_i$ and couplings $J_{ij}$ encode a particular optimization problem. Quantum annealing starts with setting $\Delta\gg\mathcal{E}$, then $\Delta$ is reduced and $\mathcal{E}$ is increased until $\mathcal{E}\gg\Delta$ and $H_S\approx H_P$.

Thus, initially the qubit register is prepared in an eigenstate of the $\sigma^x_{tot} = \sum_i \sigma_i^x$ operator and then the following evolution $\sum_{i<j} J_{ij}\,\sigma_i^z\,\sigma_j^z$ is applied (for simplicity we assume $h_i=0$). If we now take $J_{ij} = J_n$ for two neighboring qubits and $J_{ij} = 0$ otherwise, we will reproduce the evolution in the XY model, where the register is initially in the eigenstate of $\sigma^z_{tot}$ and evolution takes place in the orthogonal subspace $\sum_n \dfrac{J_n}{2} \left[\sigma_n^x\,\sigma_{n+1}^x+\sigma_n^y\,\sigma_{n+1}^y\right]$.

\subsection{How to perform the QKT of MRI data?}

The MRI frequency data form a matrix of complex coefficients $\{ f_{x,y}\}$, $x,y = 1, \dots, N$, and their processing requires a two-dimensional QKT. Thus, the input data have to be encoded in a 2D quantum superposition with $\{f_{x,y}\}$ defining its amplitudes For a spin chain implementation this could be the following encoding
\begin{align}
\ket{\Psi_{in}} = \sum_{i,j=1}^{N} f_{i,j} \ket{0_1, \dots,1_i,\dots, 0_N} \ket{0_1, \dots,1_j,\dots, 0_N}.
\end{align}
Here a long chain of spins is divided into two subchains, and the operations on them are performed independently.


\begin{thebibliography}{99}
		
		
		\bibitem{Zoller2005} P. Zoller, et al., Quantum information processing and communication: Strategic report on current status, visions and goals for research in Europe,  \textit{Eur. Phys. J. D} \textbf{36}, 203 (2005). 
		
		\bibitem{Brigham1988} E. Brigham, \textit{Fast Fourier Transform and Its Applications.} (Prentice Hall, Upper Saddle River, NJ, USA, 1988).
		
		\bibitem{Nielsen2000} M. Nielsen, I. Chuang, Quantum Computation and Quantum Information (Cambridge University Press, Cambridge, UK, 2000), pp. 216--242.
		
		\bibitem{Atakishiyev1997} N. M. Atakishiyev, K. B. Wolf, Fractional Fourier–Kravchuk transform, \textit{JOSA A} \textbf{14}, 1467 (1997). 
		
		\bibitem{Weimann2015} S. Weimann, A. Perez-Leija, M. Lebugle, R. Keil, M. Tichy, M. Gr\"afe, R. Heilmann, S. Nolte, H. Moya-Cessa, G. Weihs, D. N. Christodoulides, A.  Szameit, Implementation of quantum and classical discrete fractional Fourier transforms, \textit{Nature Commun.} \textbf{7}, 11027 (2016).
		
		\bibitem{Crespi2016} A. Crespi, R. Osellame, R. Ramponi, M. Bentivegna, F. Flamini, N. Spagnolo, N. Viggianiello, L. Innocenti, P. Mataloni, F. Sciarrino, Suppression law of quantum states in a 3D photonic fast Fourier transform chip, \textit{Nature Commun.} \textbf{7}, 10469 (2016).
		
		\bibitem{Hong1987} C. K. Hong, Z. Y. Ou, L. Mandel,  Measurement of subpicosecond time intervals between two photons by interference. \textit{Phys. Rev. Lett.} \textbf{59}, 20442046 (1987).
		
		\bibitem{Makino2016} K. Makino, Y. Hashimoto, J.-I. Yoshikawa, H. Ohdan, T. Toyama, P. van Loock, A. Furusawa, Synchronization of optical photons for quantum information processing, \textit{Science Adv.} \textbf{2}, e1501772 (2016).
		
		
		\bibitem{Sejdic2010} E. Sejdi\'c, I. Djurovi\'c, L. Stankovi\'c, Fractional Fourier transform as a signal processing tool: An overview of recent developments, \textit{J. Sig. Pro.} \textbf{91}, 1351 (2010).
		
		\bibitem{Morgenstern1973} J. Morgenstern, Note on a lower bound on the linear complexity of the fast Fourier transform, \textit{J. ACM}, \textbf{20}, 305 (1973).
		
		\bibitem{Ailon2015} N. Ailon, Tighter fourier transform lower bounds. In \textit{International Colloquium on Automata, Languages, and Programming} (Springer, 2015), pp. 14--25.
		
		\bibitem{Klappenecker} C. M. Bowden, G. Chen, Z. Diao, A.  Klappenecker, The universality of the quantum Fourier transform in forming the basis of quantum computing algorithms, \textit{J. Math. Anal. Appl.} \textbf{274}, 69 (2002). 
		
		\bibitem{Gottesman} D. Gottesman, Fault-tolerant quantum computation with higher-dimensional systems, \textit{Chaos Solitons Fractals} \textbf{10}, 1749 (1999).
		
		\bibitem{Hales2000} L. Hales, S. Hallgren, An improved quantum Fourier transform algorithm and applications, \textit{IEEE Proc. 41st Annual Symposium on Foundations of Computer Science} (2000).
		
		\bibitem{Yap2003} P. T. Yap, R. Paramesran, S. H. Ong,  Image analysis by Krawtchouk moments, \textit{IEEE Transactions on image processing}, \textbf{12}, 1367 (2003). 
		
		\bibitem{Kumar2015} A. Kumar, Nonlocal means image denoising using orthogonal moments, \textit{Applied optics}, \textbf{54}, 8156 (2015).
		
		\bibitem{Venkataramana2011} A. Venkataramana, P. A. Raj, Recursive computation of forward Krawtchouk moment transform using Clenshaw's recurrence formula, \textit{Computer Vision, Pattern Recognition, Image Processing and Graphics (NCVPRIPG), 2011 Third National Conference on. IEEE} (2011).
		
		\bibitem{Campos1989} R. A. Campos, B. E. A. Saleh, M. C. Teich, Quantum-mechanical lossless beam splitter: SU (2) symmetry and photon statistics, \textit{Phys. Rev. A} \textbf{40}, 1371 (1989).
		
		\bibitem{Kim2002} M. S. Kim, W. Son, V. Buzek, P. L. Knight, Entanglement by a beam splitter: Nonclassicality as a prerequisite for entanglement, \textit{Phys. Rev. A} \textbf{65}, 032323 (2002).
		
		\bibitem{Humphreys2015} P. C. Humphreys, B. J. Metcalf, T. Gerrits, T. Hiemstra, A. E. Lita, J. Nunn, S. W. Nam, A. Datta, W. S. Kolthammer, I. A. Walmsley, Tomography of photon-number resolving continuous-output detectors, \textit{New J. Phys.} \textbf{17}, 103044 (2015).
		
		\bibitem{Hofheinz2009} M. Hofheinz, H. Wang, M. Ansmann, R. C. Bialczak, E. Lucero, M. Neeley, A. D. O'Connell, D. Sank, J. Wenner, J. M. Martinis, A. N. Cleland, Synthesizing arbitrary quantum states in a superconducting resonator, \textit{Nature} \textbf{459}, 546 (2009). 
		
		\bibitem{Raymer2009} M. G. Raymer, S. J. van Enk, C. J. McKinstrie, H. J. McGuinness, Interference of two photons of different color, \textit{Opt. Commun.} \textbf {283}, 747--752  (2009).
		
		\bibitem{Kobayashi2016} T. Kobayashi, R. Ikuta, S. Yasui, Sh. Miki, T. Yamashita, H. Terai, T. Yamamoto, M. Koashi, N. Imoto, Frequency-domain Hong--Ou--Mandel interference, \textit{Nature Phot.} \textbf {10}, 441--444 (2016).
		
		\bibitem{DWavePrivateComm} D. Dahl, Qubits, Couplers \& Quantum Computing in 2017, ISC High Performance 2017 Conference (Frankfurt, 2017).
		
		\bibitem{Fan2012} H.-Y. Fan, L.-Y. Hu, Correspondence between quantum-optical transform and classical-optical transform explored by developing Dirac’s symbolic method, \textit{Front. Phys.} \textbf{7}, 261--310 (2012).
		
		\bibitem{Pavlidis2017} A. Pavlidis, E. Floratos, Arithmetic Circuits for Multilevel Qudits Based on Quantum Fourier Transform, preprint at https://arxiv.org/abs/1707.08834 (2017).
		
		\bibitem{Cyranoski2017} D. Cyranoski, China launches brain-imaging factory, \textit{Nature} \textbf{548}, 268 (2017).
		
		\bibitem{Eckstein2011} A. Eckstein, A. Christ, P. J. Mosley, Ch.  Silberhorn, Highly efficient single-pass source of pulsed single-mode twin beams of light, \textit{Phys. Rev. Lett.} \textbf{106}, 013603 (2011).
		
		\bibitem{Gerrits2011} T. Gerrits et al., On-chip, photon-number-resolving, telecommunication-band detectors for scalable photonic information processing, \textit{Phys. Rev. A} \textbf{84}, 060301 (2011).
		
		\bibitem{Feinsilver2005} P. Feinsilver, Philip, J. Kocik, Krawtchouk polynomials and Krawtchouk matrices. In: \textit{Recent advances in applied probability} (Springer, Boston, MA, 2005), pp. 115–141.
		
		\bibitem{Papakostas2011} G. A. Papakostas, D. E. Koulouriotis, E. G. Karakasis, V. D. Tourassis, A General Framework for Computation of Biomedical Image Moments. In: \textit{Biomedical Engineering, Trends in Electronics, Communications and Software} (INTECH, 2011).
		
		\bibitem{Gautam2017} G. Gautam, K. Choudhary, S. Chatterjee, M. H. Kolekar, Facial expression recognition using Krawtchouk moments and support vector machine classifier. In: \textit{Fourth IEEE International Conference on Image Information Processing (ICIIP)} (2017) pp. 1--6.
		
		\bibitem{Liu2017} X. Liu, G. Han, J. Wu, Z. Shao, G. Coatrieux, H. Shu, Fractional Krawtchouk transform with an application to image watermarking, \textit{IEEE Transactions on Signal Processing} \textbf{65}, 1894 (2017).
		
		\bibitem{Mademlis2006} A. Mademlis, A. Axenopoulos, P. Daras, D. Tzovaras, M. G. Strintzis, 3D content-based search based on 3D Krawtchouk moments, In: \textit{Third IEEE International Symposium on 3D Data Processing, Visualization, and Transmission (3DPVT'06)} (2006), pp. 743--749.
		
		\bibitem{Papakostas2009} G. A. Papakostas, B. G. Mertzios, D. A. Karras, Performance of the orthogonal moments in reconstructing biomedical images. In: \textit{IWSSIP 2009 – IEEE 16th International Conference on Systems, Signals and Image Processing} (2009), pp. 1--4.
		
		\bibitem{Atakishiev1994} N. M. Atakishiyev, K. B. Wolf,  Approximation on a Finite-Set of Points Through Kravchuk Functions. \textit{Rev. Mex. Fis.} \textbf{40}, 366 (1994).
		
		\bibitem{Klyshko1980} D. N. Klyshko, Use of two-photon light for absolute calibration of photoelectric detectors. \textit{Sov. J. Quantum Electron.} \textbf{10}, 1112--1117 (1980).
		
		\bibitem{Christ2011} A. Christ, K. Laiho, A. Eckstein, K. Cassemiro, Ch. Silberhorn, Probing multimode squeezing with correlation functions. \textit{New J. Phys.} \textbf{13}, 033027 (2011).
		
		\bibitem{Filip2002} R. Filip, Overlap and entanglement-witness measurements, \textit{Phys. Rev. A} \textbf{65}, 062320 (2002).

		\bibitem{Christandl2004} M. Christandl, N. Datta, A. Ekert, A. J. Landahl, Perfect State Transfer in Quantum Spin Networks, \textit{Phys. Rev. Lett.} \textbf{92}, 187902 (2004).
		
	\end{thebibliography}
\end{document}